\documentclass[11pt]{article}
\usepackage[margin=1in]{geometry}
\usepackage{amsmath}
\usepackage{booktabs}
\usepackage{array}
\newsavebox{\fitbox}
\newcommand{\fitwidth}[1]{\sbox{\fitbox}{#1}\ifdim\wd\fitbox>\textwidth\resizebox{\textwidth}{!}{\usebox{\fitbox}}\else\usebox{\fitbox}\fi}
\usepackage{graphicx}
\usepackage{float}
\usepackage{rotating}
\usepackage{subcaption}
\usepackage[hidelinks]{hyperref}
\usepackage[T1]{fontenc}
\usepackage{lmodern}
\usepackage{microtype}
\usepackage{placeins}

\title{FastSAR: Open-Source SAR Image Formation on CPUs, GPUs, and TPUs, and a Matched Comparison of Algorithms, Precisions, and Processors}
\author{Paul Singerman, PhD\\ Penn State University\\ \texttt{pgs16@psu.edu} \and Tom Braun, PhD\\ Penn State University\\ \texttt{trb5864@psu.edu}}
\date{November 2026}

\begin{document}
\maketitle

\begin{abstract}
The number of spaceborne synthetic aperture radar (SAR) collectors is growing exponentially, each returning raw collections that must be formed into images. Exact backprojection needs 1.6~trillion interpolated reads and 90 trillion floating-point operations for a 108-megapixel Umbra image. How long it takes to form an image, and at what price, depends on the algorithm, the arithmetic, and the hardware. Open-source implementations have never been compared on matched hardware against a common reference, and none run on Google tensor processing units (TPUs). This paper presents FastSAR, an open-source SAR image formation library for central processing units (CPUs), NVIDIA graphics processing units (GPUs), and TPUs. With it, we compare three SAR image formation algorithms, four floating-point formats, and six cloud devices in every supported combination, with cost per image at on-demand cloud prices as the fair common measure. FastSAR is 1.06 to 22{,}000 times faster than the tested open-source implementations at equal or better accuracy, and the full suite of hardware, algorithm, and floating-point combinations lets a user choose the Pareto-optimal configuration for their balance of accuracy, speed, and cost. The L4 GPU forms images the cheapest overall. Larger GPUs are faster but cost more per image because their price rises faster than their speed. The TPUs come close to the GPUs in cost, but only for the factorized algorithm, and the CPU is far less efficient in both speed and cost. Lower precision formats buy less of a speedup than expected, and each step down in precision costs accuracy.
\end{abstract}

\section{Introduction}

\begin{figure}[!t]
\centering
\includegraphics[width=\columnwidth]{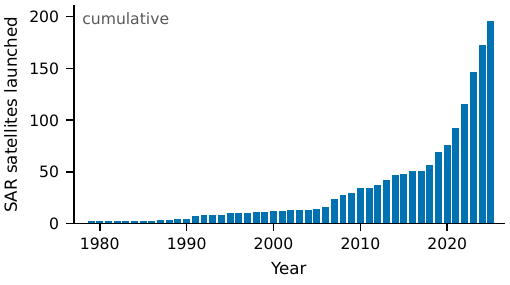}
\caption{The number of SAR satellites launched has grown exponentially since 2018. Cumulative count at the end of each year, 2025 through 4 July, read from the curve of D'Errico~\cite{derrico2025}.}
\label{fig:launches}
\end{figure}

The number of synthetic aperture radar (SAR) satellites is growing exponentially (Figure~\ref{fig:launches})~\cite{derrico2025}. The rise since 2018 comes from commercial microsatellite constellations~\cite{derrico2025,moreira2025}, which launched about 95 satellites from 2018 to 2024~\cite{moreira2025}. ICEYE alone has launched more than 60 satellites since its first in January 2018~\cite{moreira2025,eoportaliceye}, and Umbra, whose collections this paper uses~\cite{umbra2023}, launched its eleventh in 2025~\cite{eoportalumbra}. Each satellite returns raw phase histories that must be formed into images before the data can be used, and the demand for image formation grows with the fleets.

Exact backprojection forms an image without far-field or planar-wavefront approximations~\cite{jakowatz1996,gorham2010}, but it reads one interpolated sample per pixel for every pulse and multiplies each by a phase factor. For the Umbra image of the Panama Canal used throughout this paper, that is 1.6~trillion interpolated reads and about 90 trillion floating-point operations (Appendix~\ref{app:algorithms}).

How efficiently those operations, or an approximation of them, are carried out depends on which algorithm is used, in what arithmetic, and on which processor, and no published study has varied all three in one measurement. The open-source implementations that form this image span four orders of magnitude in time. Three of the backprojections run on a central processing unit (CPU) instance with 16 virtual CPUs (vCPUs) and would take an estimated 4.8 to 69~h; the fastest, NVIDIA's MatX \texttt{sar\_bp} example~\cite{matx} as published, takes 12.5~s on an NVIDIA L4 graphics processing unit (GPU) (Table~\ref{tab:oss}).

These implementations have never been run on matched hardware or scored against a common reference, so their times do not show how much of the difference is due to the algorithm, the implementation, or the hardware, nor what accuracy each reaches. Google's tensor processing units (TPUs)~\cite{jouppi2017,jouppi2023}, accelerators built for the dense matrix products of machine learning and rented by the hour like GPUs, have not been used for SAR image formation.

We present FastSAR, an open-source library that forms SAR images on CPUs, NVIDIA GPUs, and TPUs through one Python interface (Section~\ref{sec:fastsar}), and use it to measure these choices together. We compare three SAR image formation algorithms (factorized backprojection, exact backprojection, and polar format), four floating-point formats (float64, float32, float16, and bfloat16), and six cloud devices in every combination the hardware supports. The open-source implementations run on the same instances, which makes their times directly comparable with FastSAR's. Each image is scored against a float64 exact backprojection, and cost per image at on-demand cloud prices is the common measure (Section~\ref{sec:protocol}).

At matched accuracy on the same hardware, the algorithm sets the speed and the implementation adds little. Measured this way, the tested open-source implementations take 1.06 to 22{,}000 times longer than FastSAR to form the Panama image at the same accuracy or worse (Section~\ref{sec:oss}, Table~\ref{tab:oss}), and the two ends of that range show why. The low end is the pair of exact backprojections, FastSAR's and MatX's with float64 pixel coordinates, which run within 6\% of each other on the same L4. The 3.8-fold gain of FastSAR's factorized backprojection, which forms the image in 3.0~s, therefore comes from the algorithm. Its errors are $-56$ to $-63$~dB on the CPU, the L4, and the TPUs, well within the requirements of interferometry and radiometry, the most demanding uses we know of (Section~\ref{sec:quality}).

\begin{figure}[!t]
\centering
\includegraphics[width=0.82\textwidth]{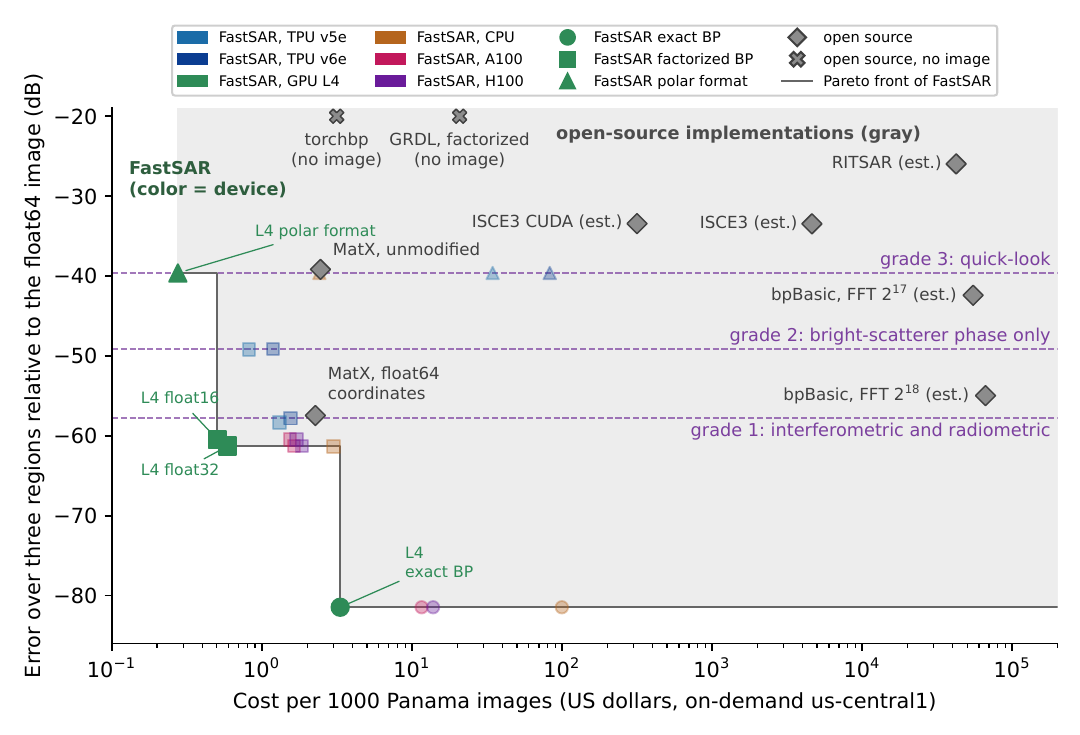}
\caption{Every open-source implementation costs more than a FastSAR configuration of equal or better accuracy. Cost per thousand Panama Canal images against the error (\ref{eq:error}) over the scored regions, where the open-source implementations were measured, for FastSAR on each device (color) and for the open-source implementations on the same CPU instance and L4 (gray; est.: estimated times). torchbp and GRDL formed no image and sit at the top edge; the line is FastSAR's cost-error Pareto front, and the dashed lines mark the image-quality grades of Section~\ref{sec:quality}.}
\label{fig:teaser}
\end{figure}

The six devices, three algorithms, and four formats map out a cost-accuracy front from which a user can pick the configuration for a given balance of accuracy, speed, and cost (Figure~\ref{fig:teaser}). The cheapest image comes from the lowest-priced GPU, the L4, at US\$0.59 per thousand. The A100 and H100 form it faster, but their hourly prices rise faster than their speed. The H100 is 5.0 times faster than the L4 at 3.1 times the cost per image.

The TPUs cost about as much per image as the larger GPUs, \$1.31 to \$1.55 per thousand, but only for the factorized algorithm, whose dense matrix products suit them. The CPU instance is the least efficient on both counts, 3.7 times slower than the L4 at a higher hourly price, and its images cost 5.1 times as much (Section~\ref{sec:cost}). The L4 stays the cheapest, within timing resolution, on seven of the eight collections measured.

Reduced precision speeds the kernels up by less than its halved data width suggests, and each format below float32 loses accuracy. Float16 in data and arithmetic on the GPUs loses about 1~dB against float32 and still supports interferometry and radiometry (Section~\ref{sec:fastsar} defines the formats), but lowers the cost by only 6 to 15\%; single-pass bfloat16 on the TPUs lowers it by 24 to 37\% and adds 12~dB of error (Section~\ref{sec:quality}). The reason is that 16-bit arithmetic speeds up the matrix products, while the other stages, bound by memory traffic or by an instruction rate that is the same in both formats, gain little (Appendix~\ref{app:kernels}).

\section{Related work}
\label{sec:prior}

Backprojection has been ported to field-programmable gate arrays (FPGAs)~\cite{cordes2009}, many-core CPUs~\cite{park2012}, and GPUs, in single-GPU CUDA implementations~\cite{fasih2010}, gigapixel formation on GPU clusters~\cite{benson2012}, nonuniform fast Fourier transform (NUFFT) interpolation~\cite{capozzoli2013}, and float16 arithmetic~\cite{xu2022}, each report optimizing one implementation for its own hardware.

The three algorithms we compare have established formulations and known accuracy limits. Fast factorized backprojection follows the quadtree of McCorkle and Rofheart~\cite{mccorkle1996} and the formulation of Ulander, Hellsten, and Stenstr\"om~\cite{ulander2003}, and has been applied to nonlinear tracks~\cite{frey2009} and to circular SAR on GPUs~\cite{ponce2011}. Polar format is described by Carrara et al.~\cite{carrara1995} and Jakowatz et al.~\cite{jakowatz1996}; Jakowatz and Doren compared it with backprojection~\cite{jakowatz2006}, and Doerry~\cite{doerry2007} and Gorham and Rigling~\cite{gorham2016} derived its scene-size limit.

No open-source implementation has been compared with another on matched hardware, and none runs on a TPU. RITSAR~\cite{ritsar}, GRDL~\cite{grdl}, and ISCE3~\cite{isce3} run on the CPU, as does the Air Force Research Laboratory (AFRL) toolbox~\cite{gorham2010}, with the backprojection \texttt{bpBasic} and the polar format \texttt{pfa\_mem}. On the GPU there are torchbp~\cite{torchbp}, ISCE3's CUDA version, and the backprojection example of NVIDIA's MatX library~\cite{matx}. Outside machine learning, the TPU has been used for Fourier transforms~\cite{lu2020,lu2021} and linear algebra~\cite{lewis2022}.

\section{FastSAR}
\label{sec:fastsar}

FastSAR takes a phase history, the antenna positions, and an output grid, and returns a complex image through one Python interface on a CPU, an NVIDIA GPU, or a TPU. The backends share the host-side float64 geometry and the factorization plan, and the devices differ only in the arithmetic they run. Around this core, the library reads vendors' phase-history and image files in the Compensated Phase History Data (CPHD), Sensor Independent Complex Data (SICD), Sentinel-1, and NASA-ISRO SAR (NISAR) formats, forms stripmap and sliding-spotlight collections as mosaics, and includes an autofocus.

The library offers three algorithms because they trade accuracy for speed in different ways (Appendix~\ref{app:algorithms}). Exact backprojection (``exact BP'' in the tables)~\cite{gorham2010} sums every pulse's contribution at every pixel and is used where accuracy outweighs time. By default it reads range profiles oversampled 8 times with cubic interpolation; at 4 times it is only a few dB more accurate than the factorized algorithm, and beyond 8 times its error no longer falls (Appendix~\ref{app:protocol}).

Factorized backprojection (``factorized BP'')~\cite{mccorkle1996,ulander2003} merges pulses into progressively longer subapertures over progressively finer image tiles, so most of the work becomes filtering and matrix products instead of per-pixel reads. It is the default because it comes within a few dB of exact backprojection at a fraction of the time (Section~\ref{sec:results}). Polar format~\cite{carrara1995,jakowatz1996} resamples the phase history onto a grid for one 2-D fast Fourier transform (FFT), the cheapest route to an image, and FastSAR follows it with a resampling that removes its planar-wavefront displacement and so improves its accuracy (Appendix~\ref{app:pfa}).

We write the factorized algorithm as tile-wise phase rotations, decimation filters, and dense matrix products, because that form runs efficiently on every device, including TPUs, whose fast units multiply dense matrices. It is implemented as one kernel per device: C++ with OpenMP for the CPU, CUDA for the GPUs, and Pallas, the kernel language of JAX~\cite{jax2018}, for the TPUs. Each stage of these kernels runs at 40 to 100\% of its hardware limit, which is memory bandwidth for the data-moving stages, arithmetic throughput for the matrix products, and sine evaluation for one level on the TPUs (Appendix~\ref{app:kernels}), so no stage is more than about 2.5 times slower than its hardware allows.

Because each accelerator's fastest arithmetic is a 16-bit format, FastSAR offers two precisions per device. On the GPUs the float32 option keeps data and arithmetic in float32 except the final matrix product. FastSAR evaluates that product on the tensor cores by splitting each operand into a bfloat16 head and a bfloat16 tail (16 significant bits in all) and summing three bfloat16 products. The float16-throughout option keeps the data and the tensor-core operands in float16 and accumulates in float32 (Appendix~\ref{app:formats}). Of the two 16-bit formats, float16 keeps 11 significant bits and bfloat16 only 8; bfloat16 gives up precision for the wider exponent range that machine learning needs~\cite{kalamkar2019}.

On the TPUs both precisions are bfloat16 and differ only in the number of passes through the matrix unit; three passes reproduce the 16 significant bits of the GPU float32 option, and one pass keeps 8. Because TPUs have no native float64 arithmetic, the TPU kernels compute the phase ramps of the last factorization level on the device in float32. We use ``float32-class'' for the arithmetic that all six devices support: float32 on the CPU, float32 with the three-pass final product on the GPUs, and three-pass bfloat16 on the TPUs.

Exact backprojection has no TPU kernel, because its one scattered read per pixel and pulse has no fast path on the TPU's vector units, which are built for dense blocks. In pure JAX it takes about 15~h per Panama image on the v5e (Appendix~\ref{app:tpu}); on TPUs, only the factorized algorithm is practical.

\section{Data and protocol}
\label{sec:protocol}

\begin{figure}[p]
\centering
\includegraphics[width=0.92\textwidth]{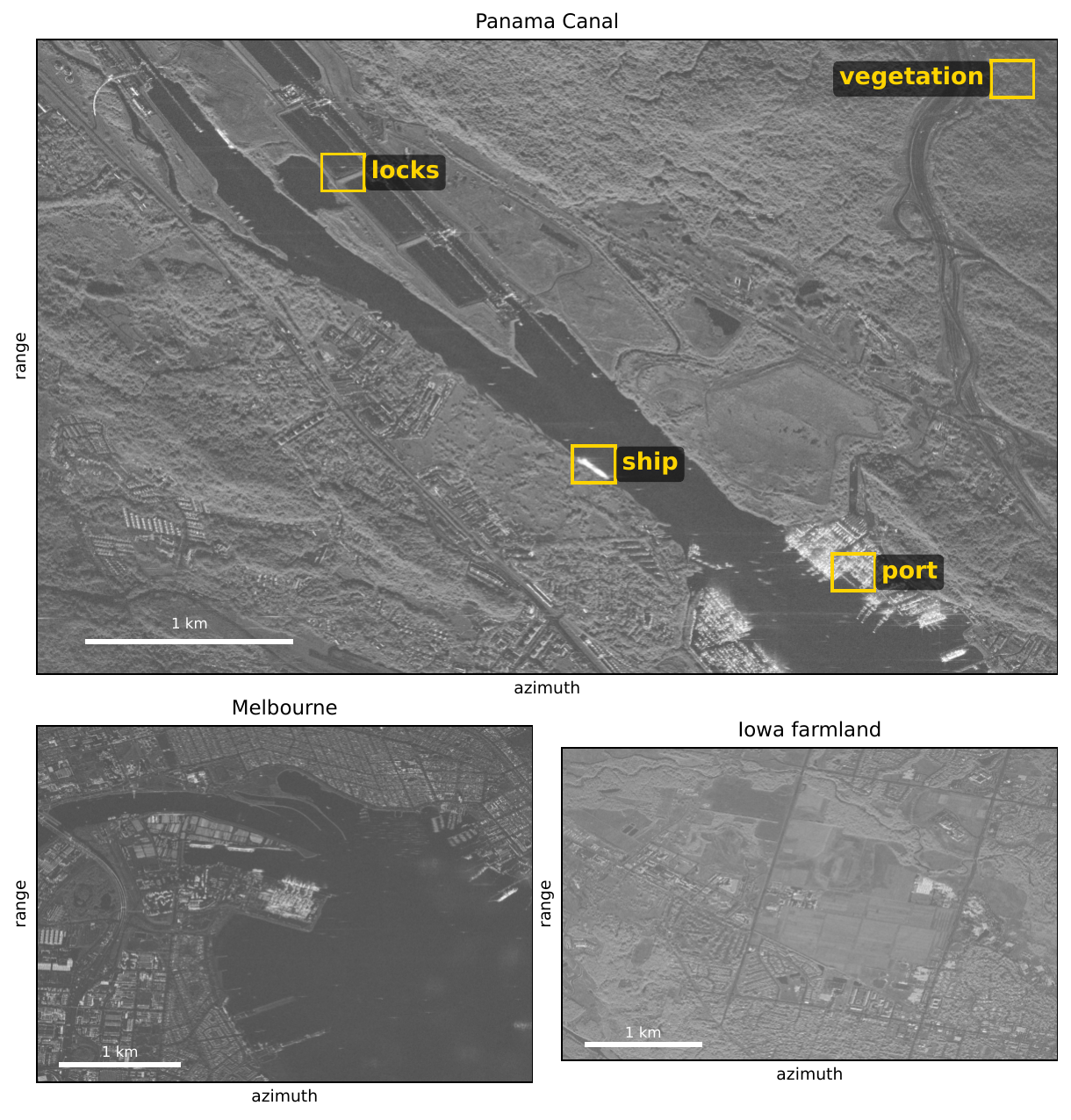}
\caption{The four Panama regions run from the scene center to a corner, so they span the planar-wavefront displacement of polar format. The three Umbra collections, with the four Panama regions used for close comparison (yellow boxes, $512 \times 512$~pixels; Figure~\ref{fig:zoom}). Formed by the float64 reference and downsampled for display (45~dB range, azimuth horizontal, range vertical, 1~km bar).}
\label{fig:scenes}
\end{figure}

We use open data from two vendors so that anyone can repeat the measurements. Three Umbra spotlight collections~\cite{umbra2023} are used throughout (Table~\ref{tab:scenes}, Figure~\ref{fig:scenes}): the Pacific entrance of the Panama Canal (July 2023), central Melbourne (February 2023), and farmland near Johnston, Iowa (October 2023). Five Capella collections~\cite{capella2023} add a second vendor and other modes: two spotlight collections (October 2024 and March 2025), a sliding-spotlight collection (September 2022; Capella's ``dynamic stripmap'' mode), and two stripmap collections (November 2021 and December 2025). Each image is formed from the CPHD file as delivered, on the vendor's image grid.

The detailed comparison uses Panama, the largest of the three Umbra images and the only strongly squinted one. It is therefore the most demanding of the three for both the approximations and the device memory, and the other collections serve as checks (Section~\ref{sec:cost}). Four of its regions, identified against Sentinel-2 imagery (Appendix~\ref{app:truth}), are examined at full resolution in Figure~\ref{fig:zoom}. Three of them, the locks, the port, and the ship (the scored regions), carry the comparison with the open-source implementations.

\begin{table}[tbp]
\centering
\caption{Panama is the largest image and the only strongly squinted one. Spacings are azimuth by range in the slant plane; squint is the Doppler cone angle less $90^\circ$ at the aperture center; range is the slant range to the scene center at the start of the aperture.}
\label{tab:scenes}
\small
\resizebox{\textwidth}{!}{\begin{tabular}{lrrrrrrr}
\toprule
Scene & Image (az $\times$ rg) & Spacing (m) & Pulses $\times$ samples & Bandwidth (MHz) & Grazing angle ($^\circ$) & Squint ($^\circ$) & Range (km) \\
\midrule
Panama Canal & 12,207 $\times$ 8,808 & 0.413 $\times$ 0.356 & 15,186 $\times$ 14,399 & 337 & 42.6 & $-28.1$ & 738 \\
Melbourne & 10,077 $\times$ 10,273 & 0.422 $\times$ 0.298 & 15,948 $\times$ 14,399 & 405 & 43.4 & 0.6 & 760 \\
Iowa farmland & 8,521 $\times$ 8,465 & 0.517 $\times$ 0.328 & 13,003 $\times$ 11,663 & 356 & 50.3 & 2.8 & 653 \\
\bottomrule
\end{tabular}}
\end{table}

All six devices are rented from Google Cloud at on-demand prices. Google Cloud is the only provider that offers TPUs~\cite{tpudocs}, and a single price list puts the devices on the same commercial footing. They are the NVIDIA L4, A100, and H100 GPUs, the TPU v5e and v6e, and a c4d-highmem-16 instance, a current general-purpose x86 instance with 8 AMD EPYC Turin cores and 16~vCPUs (the CPU instance; Table~\ref{tab:platforms}).

\begin{table}[tbp]
\centering
\caption{Hourly prices span a 16-fold range, from US\$0.71 for the L4 to US\$10.97 for the H100. Instances and on-demand prices (US dollars per hour, us-central1); the arithmetic timed on each is listed in Appendix~\ref{app:formats}.}
\label{tab:platforms}
\footnotesize
\setlength{\tabcolsep}{4pt}
\begin{tabular}{llr}
\toprule
Device & Instance & Price (\$/h) \\
\midrule
TPU v5e & v5litepod-1, 1 chip, 24-vCPU host & 1.200 \\
TPU v6e & v6e-1, 1 chip, 44-vCPU host & 2.700 \\
L4 & g2-standard-4, 1 L4 (24 GB), 4 vCPUs & 0.707 \\
A100 & a2-ultragpu-1g, 1 A100 (80 GB), 12 vCPUs & 5.02 \\
H100 & a3-highgpu-1g, 1 H100 (80 GB), 26 vCPUs & 10.97 \\
CPU & c4d-highmem-16, 8 cores, 16 vCPUs & 0.964 \\
\bottomrule
\end{tabular}
\end{table}

We score every image against a float64 exact backprojection of the same collection, because the vendor's image, formed by polar format according to its metadata, carries a planar-wavefront displacement of up to 13~pixels on Panama and cannot certify errors at the $-60$~dB level. The reference reads range profiles oversampled 16 times with linear interpolation, in float64 (Appendix~\ref{app:algorithms}). Each image is compared with it after a single fitted gain, complex for the backprojections and real for polar format, whose phase differs from the reference by a smooth screen before its final resampling (Appendix~\ref{app:protocol}). The vendor's image still gives an independent check on the reference geometry. Resampled through the planar-wavefront displacement model, it agrees with the reference in position to within 0.25~pixel root-mean-square (Appendix~\ref{app:sicd}).

The error of an image $x$ against the reference $r$ is the energy of their difference relative to that of the reference, after the fitted gain $g$:
\begin{equation}
E = 10 \log_{10} \frac{\sum_i |g\,x_i - r_i|^2}{\sum_i |r_i|^2}, \qquad g = \frac{\sum_i x_i^* r_i}{\sum_i |x_i|^2},
\label{eq:error}
\end{equation}
in dB, over the whole image unless stated otherwise (Appendix~\ref{app:protocol} defines the other statistics). FastSAR's exact backprojection is more accurate than the reference itself, whose own error against a float64 computation with profiles oversampled 64 times is $-67.3$ to $-68.8$~dB, so its region errors are measured against that computation (the 64-times computation).

The main measure is the cost per thousand images, because time alone favors the most expensive hardware and cost puts devices of very different price and speed on one scale. Time is the wall-clock time to form one image on a warm device, from the phase history in host memory to the image in host memory, with transfers and per-collection host work included. Compilation and disk reads are excluded, because a production pipeline compiles once and then forms many images. Each configuration is the library's default for its backend and precision, what a user gets without tuning.

We report the shorter of two warm calls (the single-image time), which removes one-off system noise. It agrees with the time per image of a back-to-back loop (the timed loop, Appendix~\ref{app:protocol}) within 3\% on the CPU and the L4. Differences below 3\% are therefore not resolved and are reported as equal. Cost per image is the on-demand hourly price divided by the timed loop's throughput or, where no loop was run, the price multiplied by the time.

\section{Results}
\label{sec:results}

\subsection{Comparison with open-source implementations}
\label{sec:oss}

The fastest open-source implementation trails FastSAR by the algorithm alone. MatX's \texttt{sar\_bp} example is the fastest open-source backprojection that produces an image, at 11.6~s for the Panama image on the L4 with float64 pixel coordinates (Table~\ref{tab:oss}). With the same linear interpolation, FastSAR's exact backprojection takes 10.9~s at the same error, so the two exact kernels are within 6\% of each other, and the 3.8-fold gain of FastSAR's factorized backprojection, at 3.0~s on the L4 and 11.2~s on the CPU instance, comes from the algorithm.

Against the other open-source backprojections the gap grows to two to four orders of magnitude. ISCE3 is the fastest of them (Table~\ref{tab:oss}), and FastSAR's factorized backprojection is about 1{,}500 times faster on the CPU instance and 540 times faster on the L4. RITSAR and \texttt{bpBasic}, both interpreted array implementations, are 14{,}000 to 22{,}000 times slower than FastSAR's factorized backprojection (Appendix~\ref{app:oss}).

Even FastSAR's exact backprojection is 46 times faster than ISCE3 on the CPU instance and 96 times on the L4; ISCE3's times therefore reflect its implementation, and MatX is the fair comparison for the algorithm. FastSAR is also the more accurate: in the scored regions the lowest open-source error is that of \texttt{bpBasic} with a $2^{18}$-point FFT, and FastSAR's float32 factorized image on the CPU is 4 to 7~dB more accurate in every region.

Scoring against a common reference also exposes accuracy problems that the implementations' own outputs would not reveal. The unmodified MatX example computes pixel coordinates with a float32 \texttt{linspace}. The rounding, about 0.12~mm at 2.5~km from the scene center, limits its whole-image error to $-33.9$~dB; with float64 coordinates, its region errors come within 0.2~dB of FastSAR's exact backprojection with the same interpolation. The two open-source polar-format implementations, \texttt{pfa\_mem} and GRDL's, leave the planar-wavefront displacement uncorrected, and their log-amplitude correlation with the reference is only 0.78 to 0.93, against at least 0.996 for FastSAR's polar format (Table~\ref{tab:osspfa}). ISCE3's images carry a linear phase ramp relative to the reference, a difference of phase convention rather than an error; its entries in Table~\ref{tab:oss} are measured after removing it (Appendix~\ref{app:oss}).

\begin{table}[tbp]
\centering
\caption{FastSAR's factorized backprojection is the fastest backprojection on both instances, at lower error than every open-source backprojection. Errors are the range over the scored regions, worst to best; $16\times$, $8\times$, and $2\times$ are the profile oversampling, and est.~marks times scaled from regions or patches by pixel count (Appendix~\ref{app:oss}). torchbp's 0.0 is its error in dB, an image with no correlation to the reference, and bold marks the fastest implementation on each instance with every region error below $-50$~dB, the level the best open-source backprojection reaches.}
\label{tab:oss}
\small
\fitwidth{\begin{tabular}{llrr}
\toprule
Implementation & Instance & Error (dB) & Time per full image \\
\midrule
RITSAR, as published, $16\times$ & CPU & $-25.3$ to $-27.3$ & 44~h (est.) \\
\texttt{bpBasic}, FFT $2^{17}$ (default) & CPU & $-38.1$ to $-43.2$ & 57~h (est.) \\
\texttt{bpBasic}, FFT $2^{18}$ & CPU & $-52.9$ to $-55.4$ & 69~h (est.) \\
ISCE3 \texttt{backproject} & CPU & $-31.5$ to $-54.9$ & 4.8~h (est.) \\
GRDL, factorized & CPU & no image$^\dagger$ & 77.5~s \\
FastSAR, exact, $8\times$ & CPU & $-77.5$ to $-81.7$ & 371.9~s \\
\textbf{FastSAR, factorized, float32} & \textbf{CPU} & \boldmath$-57.2$ to $-62.6$\unboldmath & \textbf{11.2~s} \\
\midrule
ISCE3 \texttt{backproject}, CUDA & L4 & $-31.5$ to $-52.2$ & 27~min (est.) \\
torchbp, exact, $2\times$ & L4 & no image$^*$ ($0.0$) & 16.0~s \\
torchbp, factorized & L4 & no image$^*$ & out of memory \\
MatX \texttt{sar\_bp}, \texttt{taylor\_fast}, $8\times$, unmodified & L4 & $-37.7$ to $-43.2$ & 12.5~s \\
MatX \texttt{sar\_bp}, \texttt{taylor\_fast}, $8\times$, float64 coordinates$^\ddagger$ & L4 & $-54.9$ to $-58.6$ & 11.6~s \\
FastSAR, exact, linear, $8\times$ & L4 & $-54.8$ to $-58.7$ & 10.9~s \\
FastSAR, exact, $8\times$ & L4 & $-77.5$ to $-81.7$ & 16.8~s \\
FastSAR, factorized, float32 & L4 & $-57.1$ to $-62.6$ & 3.0~s \\
\textbf{FastSAR, factorized, float16 throughout} & \textbf{L4} & \boldmath$-55.6$ to $-61.6$\unboldmath & \textbf{2.5~s} \\
\bottomrule
\end{tabular}}

\smallskip
\begin{minipage}{0.92\textwidth}\footnotesize
$^*$torchbp evaluates the pixel-to-antenna distance in float32, whose spacing at the 744~km mid-aperture range is 6~cm, about two wavelengths; evaluated in float64 the same formula forms the image (Appendix~\ref{app:oss}).\\
$^\dagger$GRDL's factorized backprojection is a stripmap algorithm; its image of this spotlight collection is smeared in cross-range (Appendix~\ref{app:oss}).\\
$^\ddagger$The example builds its pixel coordinates with a float32 \texttt{linspace}; this row builds them in float64 (Section~\ref{sec:oss}).
\end{minipage}
\end{table}

\subsection{Image quality}
\label{sec:quality}

Precision below float32 costs accuracy in a way a single whole-image error does not describe. Float32-class, float16-throughout, and exact images support both interferometry and radiometry. Single-pass bfloat16 keeps the phase of bright scatterers but not the amplitude of the weaker pixels around them, and polar format supports neither use. We therefore grade each image by what it can be used for. Interferometric use needs phase accuracy at bright scatterers, where persistent-scatterer interferometry measures motion to about 1~mm~\cite{ferretti2001,colesanti2003}. At the 3.1~cm wavelength of these collections, 1~mm of line-of-sight motion is about 23$^\circ$ of phase. We therefore require the phase error over the brightest 1\% of the pixels of each region to stay below a tenth of that, 2.3$^\circ$, at the 99th percentile.

Radiometric use needs amplitude accuracy well below calibration accuracy, which is a few tenths of a dB for well-calibrated systems~\cite{schwerdt2010} and about 1 to 2~dB for Umbra~\cite{csdaumbra2026}. Radiometry concerns distributed targets rather than points, so we require the amplitude ratio to the reference, over the brighter half of each region's pixels, to stay within 0.1~dB at the 99th percentile. Grade~1 meets both requirements, grade~2 the phase requirement only, and grade~3 neither (Appendix~\ref{app:protocol}; Table~\ref{tab:prec} gives the statistics).

Float32-class, float16-throughout, and exact images all reach grade~1. Their whole-image errors are $-60$~dB on the L4 and the CPU instance and $-58$~dB on the TPUs, and their region errors are $-56$ to $-63$~dB (Appendix~\ref{app:protocol}). At bright scatterers their phase error is at most 0.1$^\circ$ at the 99th percentile, less than a twentieth of the threshold, and over the brighter half their amplitude ratio is within 0.03~dB. The TPU images are about 2~dB worse than the GPU images, mainly because of the float32 geometry of their last level (Appendix~\ref{app:algorithms}).

The accuracy lost to 16-bit arithmetic is close to the rounding noise of each format, and single-pass bfloat16 concentrates its loss in the weak pixels beside bright returns. Float16 throughout on the L4 adds about 1.0~dB of error and single-pass bfloat16 on the TPUs adds 12~dB. Each is within about 1~dB of the rounding noise of the format's 11 or 8 significant bits (Appendix~\ref{app:formats}), so no implementation in that format could do better.

At the bright scatterers themselves the single-pass bfloat16 phase error stays at 0.2$^\circ$, but over the brighter half of the pixels its amplitude ratio reaches 0.16~dB at the 99th percentile, which places it in grade~2. Its worst error lies outside what the grades measure. Beside bright returns its phase error reaches 106$^\circ$ (173$^\circ$ on Melbourne).

Exact backprojection is the most accurate configuration, 19 to 21~dB below the float32-class factorized image in the scored regions. It is $-77.5$ to $-81.7$~dB from the 64-times computation there, about 23~dB more accurate than MatX, in 16.8~s on the L4.

Polar format is the only configuration in grade~3. It raises the error by 28~dB, to $-32.1$~dB, with a bright-scatterer phase error of 4.8$^\circ$ at the 99th percentile and an amplitude ratio over the brighter half of 0.69~dB. Its error grows from $-40$~dB in the inner ring to $-30$~dB in the outer ring, consistent with planar-wavefront terms beyond the first-order shift that FastSAR removes (Appendix~\ref{app:pfa}).

\begin{figure}[p]
\centering
\includegraphics[height=0.80\textheight]{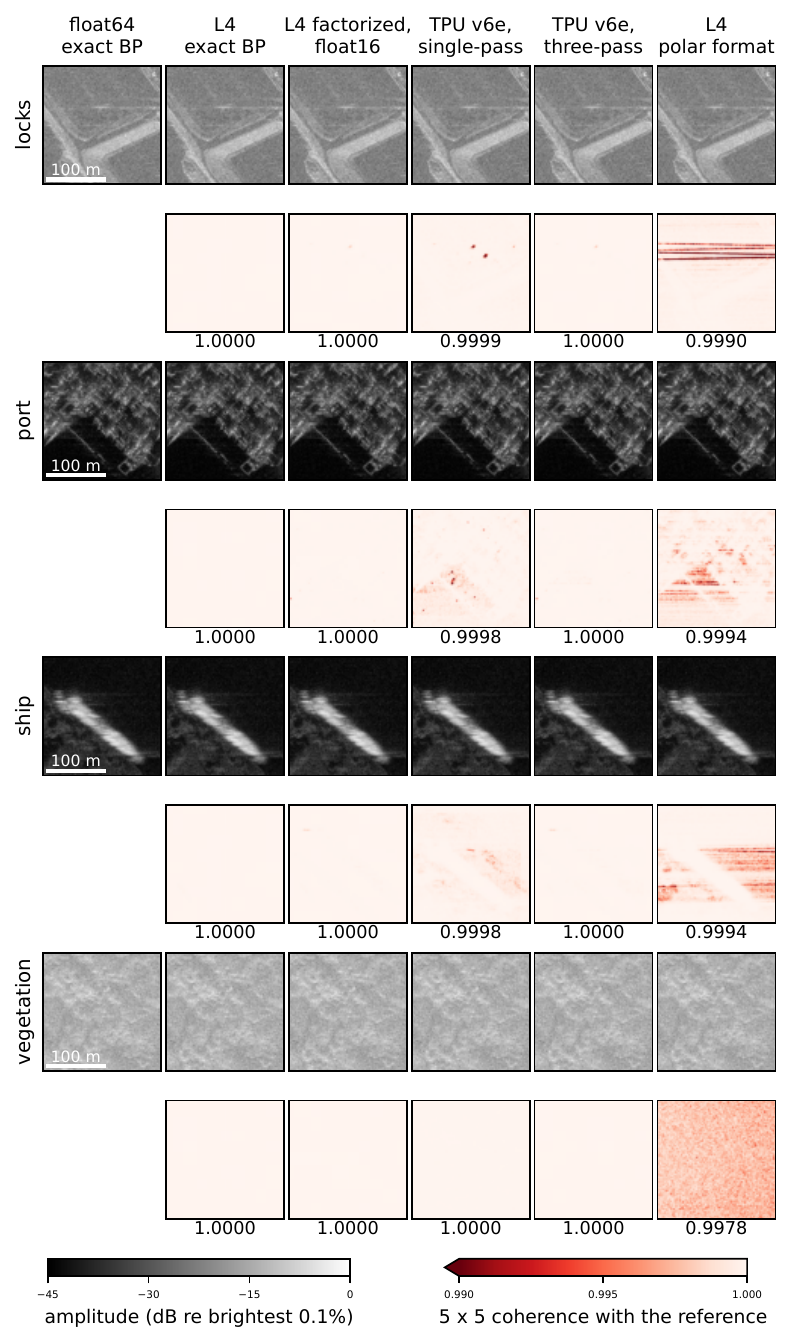}
\caption{The images are visually identical, so amplitude does not separate the methods; the coherence rows do, showing phase differences beside bright returns for single-pass bfloat16 and in stripes for polar format. Four $512 \times 512$~pixel Panama regions at full resolution (100~m bar; Appendix~\ref{app:truth}): in each pair, amplitude over 45~dB above and the $5 \times 5$ coherence with the reference below, 0.99 (red) to 1 (white), with its mean. Columns: the reference, exact backprojection, float16 throughout (L4), single-pass and three-pass bfloat16 (v6e), and polar format (L4).}
\label{fig:zoom}
\end{figure}

\subsection{Cost per image on CPUs, GPUs, and TPUs}
\label{sec:cost}

The L4 has the lowest cost per image, although three devices are faster. A thousand float32-class Panama images cost US\$0.59 on the L4 and \$1.31 to \$2.99 on the other five devices (Table~\ref{tab:cost}). The ranking by time per image is nearly the reverse, from 0.6~s on the H100 to 11.2~s on the CPU instance.

Hourly price grows more steeply than throughput as one moves up the device range. The CPU instance rents for 1.4 times the L4's hourly price yet takes 3.7 times as long; its images cost 5.1 times as much. The H100 forms an image 5.0 times faster than the L4 but its instance costs 16 times as much per hour, and a thousand images cost 3.1 times the L4's; on the A100 they cost 2.8 times.

The TPUs cost about as much per image as the larger GPUs, \$1.31 to \$1.55 per thousand against \$1.63 to \$1.84 (Table~\ref{tab:cost}), and only for the factorized algorithm. Polar format on them is 21 to 38 times slower than the factorized algorithm, and exact backprojection has no TPU kernel (Section~\ref{sec:fastsar}). The v6e forms an image faster than the L4 yet costs more per image, because its instance rents for 3.8 times the price, partly for a 44-vCPU host that is idle during formation. In the timed loop, where the TPU copies the next phase history to the device while the current image forms, the v6e takes 2.1~s per image against the L4's 3.0~s.

\begin{figure}[!t]
\centering
\includegraphics[width=0.9\textwidth]{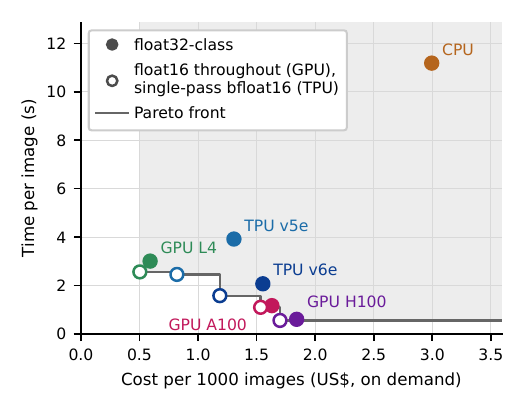}
\caption{Reduced precision never changes the cheapest device. Factorized backprojection on Panama: time per image in the timed loop against cost per thousand images, for float32-class arithmetic (filled) and for float16 throughout on the GPUs or single-pass bfloat16 on the TPUs (open). The line is the time-cost Pareto front, and the shaded area is dominated by it.}
\label{fig:speedcost}
\end{figure}

Halving the data width lowers the cost by far less than half, because only the matrix products run faster in 16-bit arithmetic (Appendix~\ref{app:kernels}). In the timed loop, float16 throughout is 1.06 to 1.17 times faster on the GPUs and lowers their cost by 6 to 15\%. Single-pass bfloat16 is 1.31 and 1.59 times faster on the v6e and v5e and lowers their cost by 24 and 37\%, with the 12~dB of added error shown in Section~\ref{sec:quality}. These reduced-precision configurations form the time-cost Pareto front (Figure~\ref{fig:speedcost}), from the L4 through the v5e and the v6e to the A100 and the H100, and the L4 stays the cheapest at either precision.

Exact backprojection costs more than the factorized algorithm, and polar format less. Exact backprojection with cubic interpolation costs \$3.32 per thousand images on the L4, 5.6 times the factorized cost. Polar format on the L4 costs \$0.28 per thousand, the lowest of any configuration, at an error of $-32$~dB. On the TPUs, polar format is 21 to 38 times slower than the factorized algorithm; its final resampling is a scattered 2-D gather of 36 reads per pixel (Appendix~\ref{app:tpu}).

The cheapest device is also the most energy-efficient. The L4 uses 0.06~kWh per thousand float32-class images, against 0.08 to 0.09~kWh on the other GPUs and the CPU instance and about 0.16 to 0.17~kWh on the TPUs (Table~\ref{tab:cost}).

Within timing resolution, the cheapest device changes on only one of the eight collections. On Melbourne and Iowa the factorized errors are within 1.2~dB of Panama's and the L4 stays the cheapest (Table~\ref{tab:costother}, Appendix~\ref{app:modes}). On the five Capella collections FastSAR's worst-region error over the five devices measured (all but the v5e) lies between $-59.1$ and $-61.5$~dB, and the H100 is the fastest device on all five. On four of them the L4 is the cheapest or within 2\% of it.

The exception is the 2025 spotlight, a 255-megapixel image whose default groups of first-level tiles do not fit in the L4's 24~GB of memory. The L4 forms it in smaller groups with more device calls, and the A100, the H100, and the v6e, whose memories hold the default groups, are cheaper on this image, the A100 by 20\% (Appendix~\ref{app:modes}).

\begin{table}[tbp]
\centering
\caption{The L4 has the lowest cost for every algorithm, and the H100 the shortest time. Every Panama configuration, cheapest first; bold rows form the cost-error Pareto front (whole-image error, ties within 0.05~dB). Time is the single-image time, cost comes from the throughput of the timed loop, and energy is that of the processor alone (Table~\ref{tab:power}); Table~\ref{tab:costother} gives Melbourne and Iowa.}
\label{tab:cost}
\small
\setlength{\tabcolsep}{4pt}
\resizebox{\textwidth}{!}{\begin{tabular}{lllrrrr}
\toprule
Device & Algorithm & Arithmetic & Time (s) & \$ per 1000 & kWh per 1000 & Error (dB) \\
\midrule
\textbf{L4} & \textbf{Polar format} & \textbf{float32, gather} & \textbf{1.4} & \textbf{0.28} & \textbf{0.02} & $\mathbf{-32.1}$ \\
\textbf{L4} & \textbf{Factorized BP} & \textbf{float16 throughout, CUDA kernels} & \textbf{2.5} & \textbf{0.50} & \textbf{0.05} & $\mathbf{-58.9}$ \\
\textbf{L4} & \textbf{Factorized BP} & \textbf{float32, CUDA kernels} & \textbf{3.0} & \textbf{0.59} & \textbf{0.06} & $\mathbf{-60.0}$ \\
TPU v5e & Factorized BP & single-pass bfloat16, TPU kernels & 3.5 & 0.82 & 0.11 & $-46.2$ \\
TPU v6e & Factorized BP & single-pass bfloat16, TPU kernels & 2.5 & 1.19 & 0.12 & $-46.2$ \\
A100 & Polar format & float32, gather & 0.9 & 1.24 & 0.03 & $-32.1$ \\
TPU v5e & Factorized BP & three-pass bfloat16, TPU kernels & 5.0 & 1.31 & 0.17 & $-57.9$ \\
A100 & Factorized BP & float16 throughout, CUDA kernels & 1.1 & 1.53 & 0.08 & $-58.9$ \\
TPU v6e & Factorized BP & three-pass bfloat16, TPU kernels & 2.9 & 1.55 & 0.16 & $-57.8$ \\
A100 & Factorized BP & float32, CUDA kernels & 1.2 & 1.63 & 0.09 & $-60.0$ \\
H100 & Polar format & float32, gather & 0.6 & 1.69 & 0.03 & $-32.1$ \\
H100 & Factorized BP & float16 throughout, CUDA kernels & 0.6 & 1.70 & 0.07 & $-58.9$ \\
H100 & Factorized BP & float32, CUDA kernels & 0.6 & 1.84 & 0.08 & $-60.0$ \\
L4 & Exact BP & linear interpolation, $8\times$ oversampled & 10.9 & 2.12 & 0.21 & $-56.8$ \\
CPU & Polar format & float32, gather & 9.0 & 2.41 & 0.06 & $-32.1$ \\
CPU & Factorized BP & float32, C++ kernels & 11.2 & 2.99 & 0.08 & $-60.0$ \\
\textbf{L4} & \textbf{Exact BP} & \textbf{cubic interpolation, $8\times$ oversampled} & \textbf{16.8} & \textbf{3.32} & \textbf{0.34} & $\mathbf{-67.4}$ \\
A100 & Exact BP & cubic interpolation, $8\times$ oversampled & 8.3 & 11.58 & 0.87 & $-67.4$ \\
H100 & Exact BP & cubic interpolation, $8\times$ oversampled & 4.5 & 13.81 & 0.71 & $-67.4$ \\
TPU v5e & Polar format & float32, gather & 103.3 & 34.44 & 4.59 & $-32.1$ \\
TPU v6e & Polar format & float32, gather & 110.1 & 82.65 & 8.42 & $-32.1$ \\
CPU & Exact BP & cubic interpolation, $8\times$ oversampled & 371.9 & 99.60* & 2.58 & $-67.4$ \\
\midrule
\multicolumn{7}{l}{Bold rows: the cost-error Pareto front. Asterisk (*): timed without a loop; cost from the single-image time. TPU energy: midpoint of the design-power band.} \\\\
\bottomrule
\end{tabular}}
\end{table}

\section{Discussion}
\label{sec:discussion}

For most uses, factorized backprojection in float32-class arithmetic on the lowest-priced GPU reaches grade~1 for less than a tenth of a cent per image, forming a 108-megapixel spotlight image in 3.0~s. Float16 throughout lowers that cost by 15\% and stays in grade~1, for users who accept 1~dB less margin. If latency dominates, the H100 is 5.0 times faster at 3.1 times the cost per image. Products that must be accurate beyond $-60$~dB, such as reference images for validating other processors, justify exact backprojection, which is about 20~dB more accurate in the scored regions for 5.6 times the time on the L4. A quick-look product that tolerates $-32$~dB can use polar format at half the cost.

Single-pass bfloat16 and polar format, the two configurations below grade~1, each fail in a specific way, and the L4's advantage has one limit. Single-pass bfloat16 on the TPUs keeps the phase of bright scatterers, but its phase errors of up to 106$^\circ$ in the weak pixels beside them would corrupt weak targets next to strong ones and the distributed scatterers that interferometry averages over. Float16 throughout on the GPUs shows no such errors. Polar format's error grows from $-40$~dB in the inner ring of the image to $-30$~dB in the outer, so its quality depends on the scene size. The L4 loses its place only when the default tile groups exceed its 24~GB of memory, as on the 255-megapixel Capella spotlight, where the A100 is 20\% cheaper.

At less than a tenth of a cent per image, the compute cost of image formation is small enough that algorithm and precision can be chosen on accuracy grounds. Several questions remain open. Spot and committed-use prices would lower every cost and could reorder devices whose discounts differ. ARM CPUs, whose price and speed differ from the x86 instance measured here, could move the CPU's place in the ranking. Forming one image across several devices would trade cost for latency in ways these single-device measurements do not show.

\section{Conclusions}
\label{sec:concl}

We have presented FastSAR, an open-source library that forms SAR images on CPUs, GPUs, and TPUs through one interface, and the first comparison of open-source SAR image formation implementations on matched hardware against a common float64 reference.

Running every implementation on the same hardware shows where the speed of SAR image formation comes from. The two fastest exact backprojection implementations agree within 6\% on the same GPU, so the 3.8-fold advantage of factorized backprojection over the fastest open-source exact implementation is a property of the algorithm, and the factorized image still meets interferometric and radiometric requirements.

The hardware sets the price of that speed. Across the cloud devices measured, hourly price rises faster than speed, and the lowest-priced GPU therefore gives the cheapest image, at US\$0.59 per thousand; a CPU instance costs 5.1 times as much per image and the fastest GPU 3.1 times. A TPU is a credible device for the factorized algorithm alone. Its images cost \$1.31 to \$1.55 per thousand, between the L4 and the larger GPUs, while exact backprojection and polar format, which gather scattered reads, have no fast path on it.

The 16-bit formats save less than their halved data width promises, and each loses accuracy. Float16 throughout saves at most 15\% at grade-1 accuracy, single-pass bfloat16 saves 24 to 37\% but introduces phase errors beside bright returns, and float64 was needed only to form the reference.

Future work will extend the comparison to more vendors and modes.

\paragraph{Acknowledgments.} The authors used Anthropic's Claude as an AI assistant in drafting and editing the text of this article. The authors designed the study, directed and checked all measurements, reviewed and edited all text, and are responsible for the content.

\paragraph{Code and data.} FastSAR is available at \url{https://github.com/saulpingerman/FastSAR} and from the Python Package Index at \url{https://pypi.org/project/fastsar/} (\texttt{pip install fastsar}); the measurements were made with release 0.1.7 (Appendix~\ref{app:protocol}). The cloud run scripts, the measurement records behind the tables, and the source of this paper are at \url{https://github.com/saulpingerman/sar-accel-study}. The radar data are from the Umbra~\cite{umbra2023} and Capella~\cite{capella2023} open data programs and the optical images from Copernicus Sentinel-2~\cite{sentinel2}; the scripts fetch them from their public sources.

\clearpage
\appendix
\section{Algorithms}
\label{app:algorithms}

This section gives the signal model, the three algorithms as FastSAR implements them, and the operation count behind the $9\times10^{13}$ floating-point operations of exact backprojection quoted in the introduction. The phase history is motion compensated to the scene reference point. For pulse $p$ with antenna position $\mathbf{a}_p$ and frequency sample $f_k$,
\begin{equation}
S_{p,k} = \sum_n \sigma_n \exp\!\left(-j \frac{4\pi f_k}{c}\, \Delta R_p(\mathbf{x}_n)\right),
\qquad
\Delta R_p(\mathbf{x}) = \lVert \mathbf{x} - \mathbf{a}_p \rVert - \lVert \mathbf{a}_p \rVert ,
\label{eq:signal}
\end{equation}
where $\sigma_n$ and $\mathbf{x}_n$ are the complex amplitude and position of scatterer $n$ and $c$ is the speed of light.

The CPHD files give the transmit and receive positions per pulse; their midpoint was used as $\mathbf{a}_p$. ISCE3~\cite{isce3} instead places the receiver at the transmit position advanced by the velocity times the delay; Appendix~\ref{app:modes} compares the two receive models. The CPHD files also give the start frequency and sample spacing per pulse; the mean values were used, which differ from the per-pulse values by at most 0.02 of a sample spacing.

A Taylor weighting ($\bar{n}=4$, $-35$~dB sidelobes)~\cite{taylor1955} was applied along both axes in every algorithm. The image grid is the vendor's: pixel $(i, j)$ lies at $(i - n_x/2)\,s_x\,\mathbf{e}_1 + (j - n_y/2)\,s_y\,\mathbf{e}_2$ from the reference point. Here $\mathbf{e}_1$ and $\mathbf{e}_2$ are the column and row unit vectors of the vendor's Sensor Independent Complex Data (SICD) product and $s_x$, $s_y$ its sample spacings.

\paragraph{Exact backprojection.} The reference range compresses each pulse by a zero-padded inverse FFT to a 16-times oversampled profile. Each pixel is the sum over pulses of the linearly interpolated profile at $\Delta R_p$ times the carrier phase $\exp(j 4\pi f_c \Delta R_p / c)$~\cite{gorham2010}. The range difference is evaluated in the cancellation-free form
\begin{equation}
\Delta R_p(\mathbf{x}) = \frac{q}{1 + \sqrt{1 + q / r_p}}, \qquad q = \frac{\lVert\mathbf{x}\rVert^2}{r_p} - 2\, \hat{\mathbf{u}}_p \cdot \mathbf{x},
\end{equation}
with $r_p = \lVert\mathbf{a}_p\rVert$ and $\hat{\mathbf{u}}_p = \mathbf{a}_p / r_p$. The reference implementation evaluates everything in float64 with Numba~\cite{lam2015} on the CPU.

FastSAR's exact backprojection (\texttt{ExactFormer}, with C++ and CUDA kernels) evaluates the range, the unit vector and the reciprocal range of each tile center in float64 once per pulse. Each pixel's range is then the center's plus a third-order expansion in its offset $\mathbf{d}$ from the center, $\mathbf{d}\cdot\hat{\mathbf{u}} + D/(2r) - (\mathbf{d}\cdot\hat{\mathbf{u}})D/(2r^2)$ with $D = \lVert\mathbf{d}\rVert^2 - (\mathbf{d}\cdot\hat{\mathbf{u}})^2$, evaluated in float32. With cubic interpolation the kernel performs about 56 floating-point operations per pixel-pulse pair (the expansion, the four interpolation weights and taps, the phase rotation and the accumulation, counting a fused multiply-add as two), about $9\times10^{13}$ for the $1.6\times10^{12}$ pairs of the Panama image. The tile, 32 by 32 pixels at orbital range, is reduced where the predicted error of the next term would exceed $3\times10^{-4}$~rad.

Each pulse is range compressed by a zero-padded FFT, and only the bins its tiles can reach are kept. The profiles are read with cubic Lagrange interpolation of 8-times oversampled profiles by default, or with linear interpolation of 8-times or, on request, 16-times oversampled profiles. On the L4 the Panama image, 1.6~trillion interpolated reads, takes 16.8~s, or 97 billion reads per second.

\paragraph{Factorized backprojection.} The image is divided recursively into tiles, as in the quadtree~\cite{mccorkle1996,xiao2000} and fast factorized~\cite{ulander2003} backprojection algorithms. At each level the phase history of a parent tile is multiplied by $\exp\{j 4\pi f_k (\lVert \mathbf{c} - \mathbf{a}_p \rVert - \lVert \mathbf{c}' - \mathbf{a}_p \rVert)/c\}$ to move its reference point from the parent center $\mathbf{c}'$ to a child center $\mathbf{c}$. The child's signal is then band limited in frequency and in pulse index in proportion to the child's size, and both axes are low-pass filtered and decimated by a Kaiser-windowed sinc~\cite{kaiser1974} with 70~dB design attenuation.

In the final stage the range to each pixel is linearized about the tile center and the kernel separates in the two image coordinates. The tile image is then a product of two matrices followed by a per-pixel quadratic phase correction, so no stage of the factorized algorithm requires an indexed read.

The image may be rectangular and of any size; each level divides the padded image side by a factor chosen from 8, 7, 6, 5, 4, 3 and 2. The decimation factors follow from the range and Doppler extents of a child tile. For the Panama image the three levels divide the azimuth side by 8, 8 and 6 and the range side by 8, 7 and 5. The first level decimates the 14,399 samples by 6 and the 15,186 pulses by 4. The final tiles are 32 pixels on a side, the default size, on all three collections, whose geometries keep the predicted expansion error below the kernel's threshold.

The pure JAX implementation of the factorized algorithm (the JAX program) remains in the library beside the device kernels of Appendix~\ref{app:kernels} that replaced it. It computes the phase ramps of the first two levels from float64 geometry on the host and those of level 2 from float32 geometry on the device, as do the TPU kernels. The CUDA and C++ kernels use float64 geometry at every level. Float32 geometry at level 1 adds an error that grows with the scene, so the first two levels keep float64.

The JAX program on the CPU, which like the TPU kernels computes the last level's ramps from float32 geometry, gives $-58.3$~dB on Panama, so that geometry accounts for about 1.7~dB of the up to 2.2~dB between the TPU and CUDA images. The ramps themselves are evaluated on the device in float32. The decimation filters are applied as dense matrix products on the TPU, as strided convolutions in the JAX program on the CPU, and as explicit fused multiply-adds in the CUDA and C++ kernels.

\section{Floating-point formats and the arithmetic of each configuration}
\label{app:formats}

Table~\ref{tab:arith} lists the arithmetic of every timed configuration step by step, and the measurements below show that the error each 16-bit format adds equals the rounding noise of its operands.

\begin{table}[tbp]
\centering
\caption{Only the products and, on the GPUs, the data in memory change type between the precisions; the ramps are float32 on every device, and the geometry is float64 except at the last level on the TPUs and in the JAX program and within the tiles of exact backprojection. Floating-point type of each step in the timed configurations. Geometry means tile centers, ranges and phase-ramp coefficients; ramps means the sines and cosines themselves; products means the operands of the matrix products, filters or sums, always accumulated in float32 or better.}
\label{tab:arith}
\footnotesize
\setlength{\tabcolsep}{3pt}
\begin{tabular}{>{\raggedright\arraybackslash}p{0.06\textwidth}>{\raggedright\arraybackslash}p{0.19\textwidth}>{\raggedright\arraybackslash}p{0.09\textwidth}>{\raggedright\arraybackslash}p{0.15\textwidth}>{\raggedright\arraybackslash}p{0.07\textwidth}>{\raggedright\arraybackslash}p{0.35\textwidth}}
\toprule
Device & Algorithm, label & Data in memory & Geometry & Ramps & Products \\
\midrule
GPUs, CPU & Exact BP, cubic or linear interpolation & float32 profiles & float64 per tile, float32 within the tile & float32 & float32 sums \\
GPUs & Factorized BP, float32, CUDA kernels & float32 & float64 & float32 & float32 fused multiply-adds in the levels; 3 bfloat16 products on the tensor cores in the final stage (float32 fused multiply-adds where its tables do not fit in shared memory) \\
GPUs & Factorized BP, float16 throughout, CUDA kernels & float16 & float64 & float32 & float16 data into float32 fused multiply-adds in the levels; float16 operands on the tensor cores in the final stage \\
TPU & Factorized BP, three-pass bfloat16, TPU kernels & float32 & float64 (2 levels), float32 & float32 & 3 bfloat16 passes (levels and final) \\
TPU & Factorized BP, single-pass bfloat16, TPU kernels & float32 & float64 (2 levels), float32 & float32 & bfloat16 operands, 1 pass (levels and final) \\
CPU & Factorized BP, float32, C++ kernels & float32 & float64 & float32 & float32 fused multiply-adds (levels and final) \\
CPU & Factorized BP, float32, JAX program & float32 & float64 (2 levels), float32 & float32 & float32 \\
all & Polar format, float32 & float32 & float64 & float32 & float32 sums over gathered samples; float32 FFTs \\
\bottomrule
\end{tabular}
\end{table}

Float16 (IEEE binary16, Table~\ref{tab:formats}) overflows above 65,504, so the phase history is scaled to unit peak before conversion. On the TPU every matrix product is formed from bfloat16~\cite{wang2019bfloat16} operands. A float32 operand is either rounded once (a single pass) or split into a bfloat16 head and tail with the cross terms accumulated (three passes); the three-pass product keeps the first 16 significant bits of each operand. A split into three parts (six passes, used by the JAX program) recovers float32 accuracy~\cite{jax2018}. The L4's tensor cores accept float16 operands, which the final stage of float16 throughout uses. The CPU kernels use no reduced-precision arithmetic; the AVX-512 bfloat16 instructions of the processor were not evaluated.

\begin{table}[tbp]
\centering
\caption{Float16 keeps 11 significant bits and bfloat16 8, an 18~dB difference in operand resolution. Floating-point formats: significant bits include the implicit bit; the spacing is the distance between adjacent representable values near 1.}
\label{tab:formats}
\begin{tabular}{lrrrl}
\toprule
Format & Bits & Significant bits & Exponent bits & Spacing near 1 \\
\midrule
float64 & 64 & 53 & 11 & $2.2 \times 10^{-16}$ \\
float32 & 32 & 24 & 8 & $1.2 \times 10^{-7}$ \\
float16 (IEEE binary16) & 16 & 11 & 5 & $9.8 \times 10^{-4}$ \\
bfloat16 & 16 & 8 & 8 & $7.8 \times 10^{-3}$ \\
\bottomrule
\end{tabular}
\end{table}

The measured image errors are consistent with the operand resolutions of Table~\ref{tab:formats}. Rounding each operand of the factorized products to bfloat16 adds noise about 48~dB below the operand~\cite{higham2002}. Once the error of the three-pass image is removed, the measured error of the single-pass TPU image agrees to within about 1~dB with the rounding noise of the two operands accumulated through three levels. The L4's float16 operands have 18~dB finer resolution than bfloat16, and their rounding adds 1.0~dB to the error of the float32 CUDA kernels, whose level products are float32 and whose final products are three-pass. A configuration in which the phase ramps are also computed in bfloat16 produces an unusable image, so the library does not offer it.

\section{Exact backprojection on the TPU}
\label{app:tpu}

FastSAR has no TPU kernel for exact backprojection because scattered per-pixel reads have no fast path on the TPU; on a TPU \texttt{ExactFormer} runs in pure JAX (linear interpolation of 32-times oversampled profiles, chosen to approach the accuracy of the cubic default). Timed on 512 by 512~pixel grids centered on the Panama lock region, with all pulses, it performed $3.0 \times 10^{7}$ interpolated reads per second on the v5e and $9.2 \times 10^{6}$ on the v6e. At those rates the 1.6~trillion reads of the Panama image take about 15~h (about \$18 per image at the instance price) and 49~h (about \$133), compared with 16.8~s on the L4. The grids agreed with the reference to $-65.9$ and $-66.0$~dB. On a 256 by 256 grid the same program ran at $2.7 \times 10^{7}$ reads per second on the v6e; the slower measurement is used above, so 49~h is an upper bound.

A TPU core has one or more matrix units, which multiply dense bfloat16 tiles at the chip's rated rate, and a vector unit~\cite{jouppi2017,tpudocs}. In microbenchmarks, data-dependent one-dimensional gathers on the vector unit run at about $10^8$ reads per second and the scattered two-dimensional gathers of the polar-format resampling at about $4 \times 10^7$, two orders of magnitude below the matrix units' rate per element. At the 36 reads per pixel of that resampling (Appendix~\ref{app:pfa}), the second rate gives the roughly 100~s polar-format time of the TPUs in Table~\ref{tab:cost}, 21 to 38 times their factorized time.

Google's Cloud TPU documentation~\cite{tpudocs} describes no texture or cache path for scattered reads. The sparse cores of the v6e, intended for embedding lookups, are documented near $10^9$ rows per second per chip~\cite{tpuv6e}, which would still need about 30~min for the Panama image. A one-hot matrix product would move the lookup onto the matrix units, but its matrix has one nonzero in several thousand entries, and the dense product would exceed the work of the factorized algorithm by a comparable factor. The closest published work computes FFTs and nonuniform FFTs, the latter for MRI reconstruction, as matrix products on the TPU~\cite{lu2020,lu2021}.

\section{Kernels for the factorized algorithm}
\label{app:kernels}

Sixteen-bit arithmetic speeds up the matrix products of the factorized algorithm and little else, so it saves less time than its halved data width suggests (Section~\ref{sec:cost}). On the TPUs single-pass bfloat16 shortens only the products on the matrix units, a minority of each stage's time (Table~\ref{tab:bounds}); the rotations on the vector unit, the sines XLA evaluates for the tables and the traffic of those tables take as long as before. On the L4 float16 storage halves the memory traffic that bounds the level kernel and the pulse filter, but the filters convert the float16 data to float32 before the fused multiply-adds, so the arithmetic bound is unchanged, and it lies close to the memory bound. The levels therefore gain little, and the float16 tensor-core final stage supplies most of the saving. Packed half2 arithmetic, which the CUDA guide requires for a higher 16-bit rate on the CUDA cores (Sec.~5.4.1 of~\cite{cudaguide}), was not used.

Each stage runs at 1.0 to 2.5 times the bound set by the unit that binds it (Table~\ref{tab:bounds}). The kernels replaced the JAX program after a stage-by-stage profile of the code that XLA, JAX's compiler, generated from it. In device time at float32-class precision the profiled kernels are 1.9 and 2.0 times faster than the JAX program on the v6e and the v5e, 3.8 times faster on the L4 and 18 times faster on the CPU (Table~\ref{tab:kernels}). Their float32 images are closer to the reference than the JAX program's because of the float64 geometry they use at every level (Appendix~\ref{app:algorithms}).

\begin{table}[tbp]
\centering
\caption{Fusing the stages and removing per-sample sines account for most of the 1.9 to 18 times gain over the JAX program. Device time per Panama image of the JAX program and of the profiled kernels (the build of Table~\ref{tab:bounds}; host time on the CPU, which has no transfers), and the released library's time from host memory to host memory (Table~\ref{tab:cost}).}
\label{tab:kernels}
\small
\resizebox{\textwidth}{!}{\begin{tabular}{llrr}
\toprule
Device & Build & 16-bit (s) & float32-class (s) \\
\midrule
TPU v6e & JAX program & 3.82 & 4.60 \\
 & Profiled kernels & 1.70 & 2.36 \\
 & Released library, host memory to host memory & 2.54 & 2.92 \\
\midrule
TPU v5e & JAX program & 6.49 & 9.29 \\
 & Profiled kernels & 2.75 & 4.57 \\
 & Released library, host memory to host memory & 3.53 & 4.99 \\
\midrule
L4 & JAX program & 11.16 & 13.69 \\
 & Profiled kernels & 2.90 & 3.59 \\
 & Released library, host memory to host memory & 2.53 & 3.00 \\
\midrule
CPU & JAX program & -- & 221.38 \\
 & Profiled kernels & -- & 12.36 \\
 & Released library, host memory to host memory & -- & 11.16 \\
\bottomrule
\end{tabular}}
\par\vspace{3pt}\parbox{\textwidth}{\footnotesize 16-bit means single-pass bfloat16 on the TPUs and float16 throughout on the L4; float32-class as defined in Section~\ref{sec:fastsar}. Device times were recorded in the development sessions and exclude the transfers and the per-collection host work, which the released library's time includes; the last row of each device is therefore not directly comparable with the two above it. The L4's float16 kernels were timed in a separate session from its profiled float32 kernels. The build-by-build record behind these rows, with the profiles, is in the repository (\texttt{results/fastsar/kernels.json}).}
\end{table}

\begin{table}[tbp]
\centering
\caption{Every stage runs within 1.0 to 2.5 times the bound of the unit that binds it. Each stage of the profiled kernels (the kernels row of Table~\ref{tab:kernels}; profiles in the repository) on the Panama image, single-pass bfloat16 on the TPUs and float32 elsewhere: the binding unit, its bound, the measured stage time, their ratio, and the measured time over the sum of all the bounds of the stage.}
\label{tab:bounds}
\small
\resizebox{\textwidth}{!}{\begin{tabular}{llrrrrr}
\toprule
Device & Stage & Binding unit & Bound (s) & Measured (s) & Ratio & To the sum \\
\midrule
TPU v6e & level 0 & vector unit & 0.15 & 0.22 & 1.4 & 0.60 \\
 & level 1 & vector unit & 0.30 & 0.30 & 1.0 & 0.52 \\
 & level 2 & sines (XLA) & 0.25 & 0.61 & 2.5 & 0.88 \\
 & final & matrix units & 0.13 & 0.29 & 2.2 & -- \\
\midrule
TPU v5e & level 0 & matrix units & 0.33 & 0.50 & 1.5 & 0.72 \\
 & level 1 & matrix units & 0.38 & 0.60 & 1.6 & 0.62 \\
 & level 2 & sines (XLA) & 0.40 & 0.88 & 2.2 & 0.78 \\
 & final & matrix units & 0.25 & 0.44 & 1.7 & -- \\
\midrule
L4 & levels (rotation and frequency filter) & memory & 0.73 & 1.36 & 1.9 & 1.08 \\
 & pulse filter & memory & 0.41 & 0.46 & 1.1 & 1.11 \\
 & final (float32) & arithmetic & 0.45 & 0.98 & 2.2 & 2.07 \\
\midrule
CPU & levels (rotation and frequency filter) & arithmetic & 4.19 & 4.87 & 1.2 & 0.79 \\
 & pulse filter & memory & 2.06 & 2.17 & 1.1 & 0.84 \\
 & final (float32) & arithmetic & 2.67 & 3.76 & 1.4 & 1.35 \\
\bottomrule
\end{tabular}}
\par\vspace{3pt}\parbox{\textwidth}{\footnotesize A bound divides the inherent work of a stage by a rate measured on the same device with microbenchmarks in the same framework (\texttt{ubench\_tpu.py} and \texttt{ubench\_gpu.py} in the repository): on the TPUs the vector unit on the rotation mix of the kernel with data resident on chip, the high-bandwidth memory copy, the matrix units at the product shapes of the kernel and sines in XLA; on the L4 and the CPU the inner mixes of the kernels on resident shared memory or cache and the memory copy rate. The sum of the bounds is the time the stage would take if none of its units overlapped; a dash marks a stage with one bound. Stage times are measured on the device, transfers excluded.}
\end{table}

In the terms of Appendix~\ref{app:algorithms}, a level splits each parent tile's phase history into those of its children, the tiles of the next level. Level 0 splits the whole collection into the first-level tiles; levels 1 and 2, which run inside one first-level tile per device call, are called the tile levels below. The plan and filters are the same in the JAX program and the kernels, and each kernel's image was checked against an exact backprojection of the same simulated scene at the same precision before it was timed.

\paragraph{Profile of the JAX program.} The profile places the three levels at the memory system and the final stage at sine evaluation. On the v6e one first-level tile of the Panama image takes 51~ms in the JAX program with single-pass bfloat16: 8.3~ms in level 0, 13.3 in level 1, 7.9 in level 2 and 21.4 in the final stage. XLA writes the rotated phase history of every child (24~bytes per sample, read back by the decimation product) at about 580~GB/s, a third of the v6e's nominal 1,640~GB/s~\cite{google2024trillium} (1,391~GB/s measured), and the children of the three levels total 63 billion samples per image.

The final stage materializes the two ramp matrices of every 32 by 32 tile ($2T$ by $P_f Q_f$ each, with $T$, $P_f$ and $Q_f$ the tile side and the pulses and frequency samples per tile) through five or six element-wise passes before the one product that uses them, and its direct form evaluates 400,000 sines and cosines per tile. The L4 profile shows the same distribution of time at the lower bandwidth of its convolution path.

\paragraph{TPU kernels (Pallas).} Two Pallas kernels replace the XLA stages. The level kernel takes a block of pulses and a window of columns of a parent into the core's vector memory and applies the phase ramps of up to eight children in turn. Each ramp is the product of a coarse table, one angle per 128-column chunk, and a fine table, one angle per column within the chunk. Both are precomputed once per pulse and child by XLA, so that no sine is evaluated inside the kernel: the Pallas documentation~\cite{pallasdocs} lists sines among its most expensive operations, and a kernel that evaluates the coarse table itself spends a third of its time on them. The fine tables are built directly rather than by successive doubling, because every intermediate table of fewer than 128 columns occupies full 128-lane tiles in TPU memory and the doubling roughly doubles the image time.

The rotated window is multiplied by the banded block of the decimation matrix on the matrix units, in one or three bfloat16 passes. At the two tile levels the pulse-axis decimation is accumulated on the matrix units across the pulse blocks in the same kernel, and neither the rotated window nor the frequency-decimated child reaches high-bandwidth memory. At level 0 the decimated child is too large for vector memory, and the pulse product is a separate banded product.

Children of one parent share each load, and the tile levels batch the parents of a tile into one launch. First-level tiles are formed in groups of four, and the phase history is read once per group. Where one block of outputs covers a whole row (the tile levels), the window is clipped to the data and read once; at level 0 consecutive output blocks overlap by the filter length.

The final kernel forms the $2T \times 2T$ product of Appendix~\ref{app:algorithms} on the matrix units, where $T = 32$ is the tile side in pixels and $P_f \times Q_f$ the pulses by frequency samples per tile. Both ramps are generated by a complex recurrence along the frequency index $q$. For a pixel offset $d$ along an image axis and the component $u_p$ of the unit vector to the antenna along that axis, the ramp is $\exp(-j 2\pi\, d u_p (a_0 + a_1 q))$. Here $a_0$ and $a_1$ are the cycles per meter at the first frequency sample and per sample step. This equals $X W^q$ with $X = \exp(-j 2\pi d u_p a_0)$ and $W = \exp(-j 2\pi d u_p a_1)$, and one tile needs four tables of sines and cosines of size $T \times P_f$ instead of one sine per sample.

With the recurrence the kernel evaluates no per-sample sines and the stage is no longer limited by sine evaluation, whereas a version with one sine per sample is no faster than the XLA final stage. The kernel takes one tile per step: stacking four tiles into one 256 by 256 product gains 3\% on the v6e with three-pass bfloat16 only and loses 2 to 4\% on the v5e, whose arrays are 128 by 128.

\paragraph{Unit rates.} The bounds of Table~\ref{tab:bounds} rest on unit rates measured on each device with microbenchmarks in the same framework. On the v6e, high-bandwidth memory copies run at 1,391~GB/s. The vector unit executes the level kernel's rotation mix on data resident in vector memory at 117 billion samples per second and a plain copy through vector memory at 155 billion, so the rotation runs within a factor of 1.3 of the vector load and store path. XLA evaluates 57 billion sine and cosine pairs per second, and the matrix units reach 457~TFLOP/s on a large bfloat16 product and 42~TFLOP/s at the final stage's 64-row shape. On the v5e the matrix units are about three times slower than the v6e's on a large product, against a vector unit only 12\% slower.

The L4's rates are a memory copy at 246~GB/s, the level kernel's inner mix on resident shared memory at 20.2 billion filter outputs per second, the final kernel's inner mix at 6.1 trillion fused multiply-adds per second (41\% of the chip's peak), and coalesced stores at 252~GB/s against 69~GB/s for strided stores. The CPU instance copies memory at 49~GB/s and runs the filter's inner loop at 314 billion fused multiply-adds per second on resident data and the final product's inner loop at 1.04 trillion, near the 8 cores' float32 peak of two 512-bit fused multiply-adds per core and cycle at the measured clock.

\paragraph{Overlap and residual inefficiency.} Every stage lies between its largest single bound and the sum of its bounds, which indicates partial overlap of its units (Table~\ref{tab:bounds}). The excess over the largest bound lies in table generation. In the final stage on every device the profiles place it in the ramp recurrence, which the microbenchmarks do not isolate. At level 2 on the TPUs it lies in the sines, whose rows of 352 samples need 256 sines per child and pulse for 512 rotated samples. On the v5e the matrix units, three times slower than the v6e's against a vector unit only 12\% slower, bind levels 0 and 1. The largest inefficiency identified within the factorization is the level-2 table generation on the TPUs, which angle-addition tables could reduce by about a factor of eight. That is about 0.2~s of the 1.7~s device time of the single-pass v6e image; the other TPU stages are at most 2.2 times their largest single bound and, where a sum is defined, below the sum of their bounds.

On the L4 the level kernel runs at 1.1 times the sum of its arithmetic and memory bounds, because the two overlap only partly at two blocks per multiprocessor, and the pulse filter within 1.1 times its memory bound. On the CPU the level kernel runs at 1.2 times its arithmetic bound and the pulse filter at 1.05 times its memory bound. On the L4 the remaining structural cost of the float32 kernels is the 172~GB per image that the level kernel writes and the pulse filter reads back, of which a streaming fusion of the two could remove about half, about 0.3~s of the 2.8~s of stage time in Table~\ref{tab:bounds}.

\paragraph{GPU kernels (CUDA).} On the GPUs the two kernels are written in CUDA and compiled through CuPy~\cite{okuta2017}. The level kernel loads the window of 32 pulses into shared memory with consecutive threads on consecutive columns, so that the global loads are coalesced. A pulse-major mapping of the threads would remove the bank conflicts from the filter reads but scatter the global loads and stores, which doubles the kernel's time. For each child it rotates the window from the raw samples, which each thread keeps in registers, generating the rotation of its samples as a geometric sequence. Rotating the window in place from one child to the next would make a child's rounding depend on which children were launched with it. Each thread then computes two adjacent outputs of one pulse with the 44-tap decimation filter as fused multiply-adds. For this phase consecutive threads are consecutive pulses, so no two threads of a group read the same shared-memory bank. The outputs are staged through shared memory and written row by row, and the pulse-axis filter is tiled the same way.

A Triton version of the level kernel through Pallas, limited to 16-column windows by its shared-memory limit, gains about 4\% with float16 operands and is 37\% slower in float32, so the library uses the CUDA kernels.

The final kernel builds the two ramp tables of a tile in shared memory by the same recurrence as the TPU kernel. A table entry's phase is linear in the pixel's offset along the tile, so each column of a table is built by a geometric recurrence, one complex multiplication per entry, and each phase-history sample is read once rather than once per pixel row. The CUDA-core version of the product is bound by the latency of that recurrence and by one shared-memory word per fused multiply-add; a 4 by 2 block of pixels per thread and interleaved complex tables relieve both. On GPUs of compute capability 8.0 or higher (the L4, A100 and H100) the product runs on the tensor cores: twelve warps of a block build the tables of the next pulses in shared memory while four multiply the current ones, alternating between two buffers.

For the float32 configuration each table entry is split into a bfloat16 value and a bfloat16 remainder, and three of the four cross products are accumulated in float32, the three-pass scheme of the TPU kernels (Appendix~\ref{app:formats}). In the library's tests its images agree with those of the CUDA-core float32 kernel to $-87$ to $-90$~dB. Where the double-buffered tables of a tile do not fit in shared memory, as at the final-level sample counts of the Capella mosaics on the L4, the float32 configuration uses the CUDA-core kernel.

The float16-throughout configuration takes one tensor-core product of float16 tables and stores the phase history, every intermediate and the final-stage operands in float16, accumulating in float32 inside the filters, which halves the memory traffic that bounds the level kernel and the pulse filter. It forms the Panama image in 2.5~s against 3.0~s for the float32 configuration (Table~\ref{tab:cost}), a saving of 0.5~s, less than halving a purely memory-bound stage would give, for 1.0~dB more error. Further gains from float16 in the levels would need the level filters rewritten as matrix products for the tensor cores, as in the TPU kernel.

\paragraph{CPU kernels (C++).} The level filter, the pulse filter and the final stage are written in C++ with OpenMP for the CPU instance, compiled by GCC 13 for AVX-512 with 512-bit vectors preferred, and called from the Python driver through a shared library. The host side (plan, filters, float64 geometry) is the code used for the other devices. The level kernel assigns the children of a parent to vector lanes, because a polyphase filter vectorized along the output columns needs stride-$D$ gathers that cost about 12 instructions per useful multiply-add. A sample of the parent row is then a scalar broadcast, its rotation one complex multiply per 16 children with a per-lane phase step, and the filter accumulates eight outputs per child vector in registers and requires no strided reads. The intermediate buffers are reused from call to call, which removes one page fault per 4~KB of fresh allocation inside the parallel regions.

The pulse filter reads each input row once per chunk of 16 output rows. The final stage builds its two tables three pulses at a time, by a geometric recurrence over the pulse index and another over the pixel index (four sines per pulse and table), so that the tables stay in the second-level cache while the product consumes them. The kernels run on 8 threads, one per core; 16 threads give the same time. The final stage runs at 1.4 times its arithmetic bound, with the excess in table building (Table~\ref{tab:bounds}).

\section{Polar format on an orbital aperture}
\label{app:pfa}

FastSAR's polar format reaches $-32.1$~dB on Panama because a final resampling removes the planar-wavefront displacement; this section gives the steps, the displacement model and the residual that remains. The algorithm consists of three resampling steps and a two-dimensional FFT, each expressed with FFTs, element-wise products or a single matrix product so that the same code runs on all devices; C++ and CUDA kernels implement the two wavenumber resamplings and the final gather on the CPU and the GPUs. The pulses are first resampled so that the tangent of the look angle in the image plane is uniform in pulse index, which an orbit does not give exactly. This is a 16-tap Kaiser-windowed sinc interpolation along the pulse axis, applied as a sum over 16 gathered rows on every device. Each pulse is then resampled onto a uniform radial wavenumber grid, and each wavenumber row onto a uniform cross-range grid.

The look angle is measured from the grid's range axis $\mathbf{e}_2$, so the grid must have $\mathbf{e}_2$ along the mid-aperture line of sight in the image plane, as the vendors' grids do (the Panama grid is aligned to 0.01$^\circ$). On a squinted Umbra spotlight a grid rotated 15$^\circ$ from it, with its axes along and across the track, gives $-5$~dB against exact backprojection where the aligned grid gives $-27$~dB, and the library warns beyond 2$^\circ$.

On the CPU and the GPUs the two wavenumber resamplings are direct Kaiser-windowed sinc interpolations of the windowed profiles, with the kernel read from a table. On the TPUs they are chirp-z transforms~\cite{rabiner1969}, which evaluate the same band-limited interpolation on a scaled grid with FFTs and chirps, with the outputs that would wrap around the band zeroed. The two agree to $-61$~dB on Panama.

The frequency samples are weighted by $f_c / f_k$ before resampling. Backprojection's uniform weighting of the samples corresponds to a density of $1/k$ in the wavenumber plane (the polar Jacobian) that the uniform spectral weighting of polar format omits; omitting this weighting worsens the error by about 1.5~dB.

Because the grazing angle changes along the aperture, the radial band of each pulse projected onto the image plane drifts from pulse to pulse. The raster covers the union of the bands and so keeps the full spectral support of backprojection, a trapezoid, rather than the inscribed rectangle. For these collections the inscribed rectangle would keep about half the bandwidth in the ground plane and nearly all of it in the vendor's slant plane. The wavenumber grid spacings are set by the native pixel spacings and FFT sizes of 13,824 by 10,800, which include a 300~m guard band on every side of the scene.

The range and azimuth profiles are tapered to zero within the guard, because the periodic transforms otherwise wrap echoes from beyond the image into its edges; without the taper the edge error is $-5$~dB. The range taper's scene edge and guard are projected onto each pulse's line of sight, because the profiles are in slant range and a taper placed in ground range lets returns from far outside the scene alias into the raster at steep incidence.

\paragraph{Distortion.} Polar format assumes a planar wavefront, and a scatterer at distance $\rho$ from the scene center is then focused about $\rho^2/2R$ from its position in range~\cite{carrara1995,doerry2007}. At $R = 744$~km (mid-aperture) this is 4~m at 2.5~km, about 12 pixels. The azimuth displacement is larger and depends on the squint (7.6~m in azimuth and 4.3~m in range at the Panama corners). The planar model writes the range history of a scatterer at $\mathbf{p}$ as $-\hat{\mathbf{u}}_p \cdot \mathbf{q} + c$. A least-squares fit of $\mathbf{q}$ and $c$ over the aperture to the exact history $\lVert \mathbf{p} - \mathbf{a}_p \rVert - \lVert \mathbf{a}_p \rVert$ gives the position at which polar format focuses the scatterer, $(q_x, q_y + c)$.

The fit residual is below 1~mrad of phase everywhere in the three scenes, so the image is displaced without defocus, by up to 18 pixels in azimuth and 12 in range at the Panama corners. Measured target positions agree with the model to 0.2~pixel. The mapping is fitted by cubic polynomials over the image (residual below $10^{-4}$~pixel). Before the final resampling, the 5 by 5 coherence with the reference is 0.42.

\paragraph{Final resampling.} The spectrum is placed in a raster of twice the pixel density and transformed without the carrier terms. The result is interpolated at the apparent positions with a six-tap Lanczos kernel along each axis and multiplied by the carriers evaluated at the apparent positions. Before the resampling a center crop shows a quadratic phase of about 12~rad at 211~m, which is the carrier of the range displacement; the carrier multiplication removes it, and no separate phase correction is applied. The oversampled raster is processed in blocks: the range transform is done once, and the azimuth transform and the resampling proceed in blocks of 512 output range columns. The polar-format figures in this paper are for the resampled image.

\begin{figure}[tbp]
\centering
\includegraphics[width=0.72\textwidth]{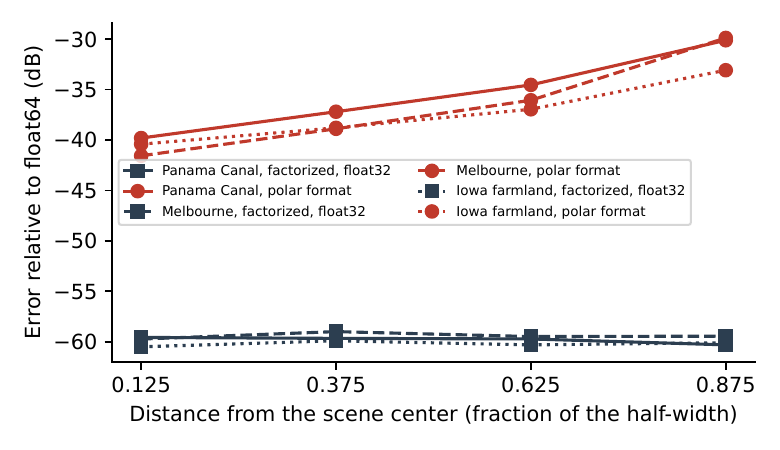}
\caption{Polar-format error grows 7 to 12~dB from the inner to the outer ring on all three collections; factorized error does not depend on distance from the center. Error relative to the reference in four concentric rings by distance from the scene center (dB; rings as in Table~\ref{tab:pfaedge}), for the polar-format image and the float32 factorized image on the L4.}
\label{fig:rings}
\end{figure}

\paragraph{Residual.} After the resampling the error is $-32.1$~dB on Panama ($-33.5$~dB on Melbourne and $-35.4$~dB on Iowa) and the mean 5 by 5 coherence 0.9994. On each of the three collections the residual grows from the inner to the outer ring, by 7 to 12~dB, from $-40$ to $-30$~dB on Panama, whereas the factorized error shows no dependence on distance from the center (Figure~\ref{fig:rings}). The displacement model removes the first-order effect of the planar-wavefront approximation, a shift of each scatterer, and fits the phase to 1~mrad at the sampled points. Terms of the approximation beyond that shift, which were not modeled, would make the residual grow with distance from the scene center.

On the full Panama image, the resampling kernel (4, 6 and 8 taps) and the oversampling of the intermediate raster (2 and 3 times) change the whole-image error by at most 0.6~dB (Table~\ref{tab:pfaedge}), and in the outermost ring, where the displacement is 10 to 18~pixels, by at most 0.3~dB. Lengthening the pulse interpolator from 16 to 32 or 48 taps, with a narrower Kaiser window, makes the error up to 0.9~dB worse. The registration error, measured from the cross-spectrum phase, is 0.001~pixel, and the convergence of the reference is more than 30~dB below the residual. The Taylor weighting, which backprojection applies per pulse and sample and polar format through the resampling onto the rectangular raster, was not tested as a cause.

\begin{table}[tbp]
\centering
\caption{No variation of the final resampling changes the polar-format error by more than 0.6~dB, and a longer pulse interpolator makes it up to 0.9~dB worse. Panama Canal, polar format on the CPU: whole-image error and error by ring (dB) for variations of the final resampling; rings are the quarters of the half-width from the scene center outward, and the selected configuration is the first row.}
\label{tab:pfaedge}
\small
\begin{tabular}{lrrrrr}
\toprule
Variation & Error & Ring 1 & Ring 2 & Ring 3 & Ring 4 \\
\midrule
6-tap kernel, 2 times oversampled (selected) & $-32.1$ & $-39.8$ & $-37.2$ & $-34.5$ & $-30.1$ \\
4-tap kernel & $-31.6$ & $-37.1$ & $-35.1$ & $-33.5$ & $-29.8$ \\
8-tap kernel & $-32.2$ & $-40.1$ & $-37.4$ & $-34.7$ & $-30.2$ \\
3 times oversampled & $-32.2$ & $-40.1$ & $-37.3$ & $-34.7$ & $-30.1$ \\
32-tap pulse interpolator, Kaiser 12 & $-31.7$ & $-37.7$ & $-35.7$ & $-33.8$ & $-29.8$ \\
48-tap pulse interpolator, Kaiser 12 & $-31.2$ & $-36.4$ & $-34.5$ & $-33.1$ & $-29.6$ \\
\bottomrule
\end{tabular}
\end{table}

\paragraph{Time.} The final resampling is an indexed read at a computed position, 36 reads per pixel, whose rate on the TPUs is given in Appendix~\ref{app:tpu}. On Panama the polar-format image takes 1.4~s on the L4 and 9.0~s on the CPU, 0.47 and 0.80 times the factorized float32 time, and 103.3~s on the v5e and 110.1~s on the v6e (Section~\ref{sec:cost}). A separable approximation of the final resampling, expressible as row and column phase ramps, would avoid the scattered gathers. The squinted geometry of these collections is not exactly separable in this way, and the approximation's effect on the TPU's cost was not measured.

\section{Validation against the vendor's image}
\label{app:sicd}

\begin{figure}[tbp]
\centering
\includegraphics[width=0.94\textwidth]{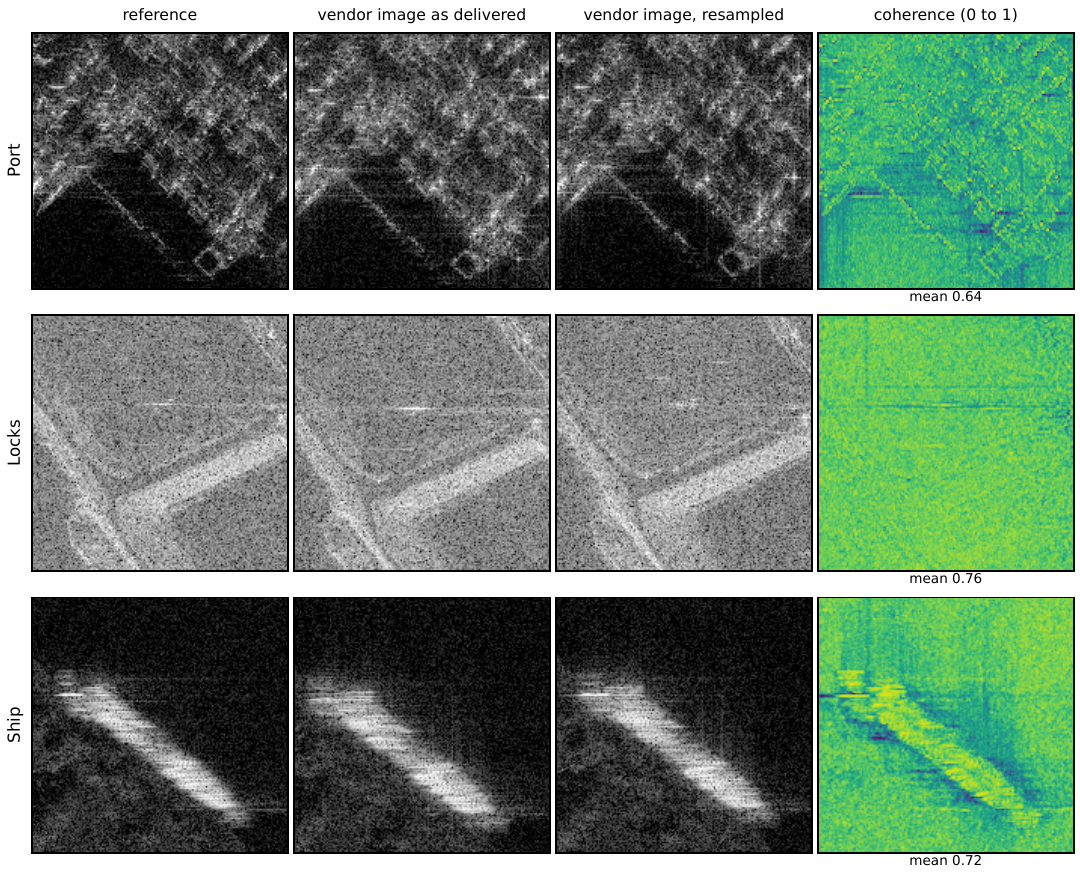}
\caption{The vendor's polar-format image is displaced by up to 13.2 pixels and, once resampled through the planar-wavefront model, registers to the reference to 0.07~pixel RMS. The port, lock and ship regions of Figure~\ref{fig:zoom} (rows, 512 by 512 pixels; the row order differs from that figure): the reference; the vendor image as delivered, on the same pixels; the vendor image after resampling through the displacement model of Appendix~\ref{app:pfa}; and the 5 by 5 coherence of the resampled vendor image with the reference after local realignment and phase-ramp removal, from 0 to 1, with its mean below each map. Amplitudes are displayed over a 45~dB range.}
\label{fig:sicd}
\end{figure}

The vendor's polar-format image is displaced from the reference by the planar-wavefront field that the collection geometry predicts, to 0.25~pixel RMS after resampling through that field, so an independent processor confirms the reference's geometry. Umbra distributes each collection as a complex image in the SICD format, on the pixel grid that the reference adopts; its metadata states that the image was formed from the same phase history by polar format. The integer and subpixel offset between the two images (consistent with the vendor's reference-pixel convention) was found from the cross-correlation of block-averaged log amplitude and refined on a central window. The linear phase ramp between them, which reflects the different spectral centering of the two images ($-0.225$ cycles per pixel in range on Panama), was found from the peak of the cross-spectrum of the same window and removed.

The vendor image was compared with the reference in 512 by 512 windows centered on a grid with 1024-pixel spacing, keeping the windows with enough contrast for an amplitude cross-correlation (68 on Panama). In each window the local offset of the vendor image was measured by amplitude cross-correlation and compared with the displacement that the planar-wavefront model of Appendix~\ref{app:pfa} predicts at the window center. The vendor image was then resampled through the model with a cubic kernel and, after a local realignment and phase-ramp removal in each window, the coherence and the log-amplitude correlation of the two images were computed. As delivered, the vendor image is masked to the parallelogram over which the vendor's processor has full spectral support, and its bright points carry cross-shaped sidelobes that the reference's Taylor weighting suppresses.

\begin{table}[tbp]
\centering
\caption{The vendor's image sits where the planar-wavefront model predicts, to 0.25 pixel RMS. The reference (float64 exact backprojection) against the vendor's complex image (SICD) of the same collection, which Umbra forms by polar format: the windows compared, the measured and the predicted displacement of the vendor image, the residual after resampling through the model, and the coherence and log-amplitude correlation after it (columns defined in the note).}
\label{tab:sicd}
\small
\setlength{\tabcolsep}{4pt}
\resizebox{\textwidth}{!}{\begin{tabular}{lrrrrrrrr}
\toprule
Collection & Windows & Measured (px) & Model (px) & RMS difference (px) & Residual after resampling (px) & Coherence, scene mean & window median & Log-amplitude correlation \\
\midrule
Panama Canal & 68 & 13.2 / 9.0 & 13.2 / 9.1 & 0.11 / 0.12 & 0.07 / 0.05 & 0.46 & 0.67 & 0.49 \\
Melbourne & 81 & 7.3 / 7.1 & 7.4 / 7.2 & 0.10 / 0.19 & 0.05 / 0.11 & 0.67 & 0.76 & 0.52 \\
Iowa farmland & 64 & 6.7 / 8.1 & 6.7 / 8.5 & 0.07 / 0.34 & 0.01 / 0.25 & 0.63 & 0.80 & 0.57 \\
\bottomrule
\end{tabular}}
\par\vspace{3pt}\parbox{\textwidth}{\footnotesize Windows: 512 by 512~pixel windows centered on a grid with 1024-pixel spacing that have enough contrast for an amplitude cross-correlation. Measured: the largest displacement of the vendor image relative to the reference over the windows (azimuth / range, pixels); Model: the largest displacement the planar-wavefront model of Appendix~\ref{app:pfa} predicts at the same windows. RMS difference: the root-mean-square difference between measured and predicted displacement; Residual: the root-mean-square displacement left after the vendor image is resampled through the model. Coherence: after that resampling and a local realignment and phase-ramp removal in each window, the 5 by 5 coherence of the two images as the mean over all pixels and as the median of the window means; Log-amplitude correlation: the median over the windows of the correlation of their log amplitudes.}
\end{table}

The vendor image (Table~\ref{tab:sicd}) is displaced from the reference by a field whose largest value over the Panama windows is 13.2~pixels in azimuth and 9.0 in range. The model predicts 13.2 and 9.1 from the collection geometry alone. The root-mean-square difference between measured and predicted displacement over the windows is 0.11 and 0.12~pixel on Panama and at most 0.34~pixel on the other two collections, 1 to 4\% of the largest displacement. After the resampling the root-mean-square residual displacement is at most 0.25~pixel (individual windows up to 0.43). The coherence of the two images averages 0.46, 0.67 and 0.63 on Panama, Melbourne and Iowa (0.67, 0.76 and 0.80 as the median over windows), and their log amplitudes correlate at 0.49 to 0.57 within windows (Figure~\ref{fig:sicd}).

At the brightest isolated points of the lock and port regions (the ship is a moving target and the vegetated region has no isolated point) the impulse-response widths of the two images differ by 1 to 27\%. The vendor's response is wider in azimuth at both points and, in range, narrower at one and 1.3\% wider at the other. The SICD metadata gives neither the aperture weighting nor whether autofocus was applied, and the vendor's motion compensation is independent of the one used here. Coherence at this level reflects mainly the differences between the two processors and is not used as a measure of accuracy. The vendor image is delivered without a correction for the displacement; the polar-format images of this study remove it in the final resampling (Appendix~\ref{app:pfa}).

The Capella 2021 stripmap and 2024 spotlight of Appendix~\ref{app:modes} were compared with the vendor's image on the vendor's pixel grid by the same method. The correlation of linear amplitudes is 0.993 for the stripmap mosaic (165 patches, 512 by 512~pixel crop) and 0.78 for the spotlight formed by exact backprojection (128 by 128 pixels at the scene center). Neither figure is comparable with the windowed log-amplitude correlation of Table~\ref{tab:sicd}. In both of these CPHD files the phase sign declared in the metadata is opposite to the sign with which the data focus, and FastSAR's reader reverses it. The sliding spotlight's frequency spacing of 114.3~kHz gives an unambiguous range of 1,311~m, shorter than its 2,962~m delay window; FastSAR does not model this convention, and its image matches the vendor's only near the scene center.

\section{Open-source implementations and their settings}
\label{app:oss}

This section gives the version, settings and scoring of each open-source implementation in Table~\ref{tab:oss} and the cause of each one's error.

\paragraph{Scoring of Table~\ref{tab:oss}.} FastSAR's region errors are measured against the 64-times computation and the open-source implementations' against the reference (Section~\ref{sec:protocol}). FastSAR's exact backprojection interpolates the oversampled profiles cubically, the others linearly; the linear FastSAR row, which exists for the comparison with MatX, is scored against the reference like the open-source rows. The float64 pixel coordinates make the MatX example faster as well as more accurate: its backprojection kernels take 10.7~s against 11.7~s unmodified, summed over the example's 60 pulse blocks; the cause was not investigated.

The full-image times marked as estimates in Table~\ref{tab:oss} scale measured region times by pixel count. RITSAR took 384~s and \texttt{bpBasic} 499~s (FFT $2^{17}$) and 604~s ($2^{18}$) on the 512 by 512~pixel lock region with all 15,186 pulses, and ISCE3 10.5~s per 256 by 256~pixel patch on the CPU instance; the CUDA version took 15.7~s for a 1024 by 1024~pixel patch on the L4. The full image has 410 times the pixels of the region, 1{,}641 times those of the CPU patch and 103 times those of the CUDA patch.

RITSAR~\cite{ritsar} (commit 0e36d2e, ported to Python~3) is a NumPy library. \texttt{bpBasic} is the backprojection of the AFRL toolbox of Gorham and Moore~\cite{gorham2010}, distributed by the National Geospatial-Intelligence Agency (commit 63f0401); it was run in GNU Octave 8.4. ISCE3~\cite{isce3} (release 0.25.17; release 0.25.21 for the Capella sliding spotlight) has C++/OpenMP and CUDA versions, and torchbp~\cite{torchbp} (commit b371884) is a PyTorch extension with CUDA kernels. GRDL~\cite{grdl} (commit de3338a) is a Python library with Numba kernels that reads CPHD files and implements polar format and fast factorized backprojection, run with its default parameters.

MatX~\cite{matx} (commit 3f3cc13) is NVIDIA's C++20 numerical library. Its \texttt{sar\_bp} example (directory \texttt{sarbp}) evaluates the exact sum with a second-order expansion of the range within each tile of pixels, as FastSAR's exact kernel does with a third-order one (Appendix~\ref{app:algorithms}), and reads 8-times oversampled range profiles with linear interpolation. It was built with CUDA 12 and run as its own executable; a converter in the study repository wrote the phase history, antenna positions and grid of the study to its input format, and its output image was scored on the three regions like the others. Its four precision modes were run; the two that keep the coordinates in float32 inside the kernel are unchanged by the float64 \texttt{linspace} (Section~\ref{sec:oss}). The float32 \texttt{linspace} rounds the coordinates by up to 0.12~mm at 2.5~km from the scene center, a two-way phase error of about 0.05~rad at the collection's 3.1~cm wavelength, consistent with the unmodified example's errors.

All implementations received the Taylor-weighted phase history and antenna positions of the reference (Appendix~\ref{app:algorithms}) in their own input formats and were scored with the error measure of Section~\ref{sec:protocol} on three of the regions of Figure~\ref{fig:zoom}, ISCE3 on 256 by 256~pixel patches of its own radar grid. The whole instance was available to every run; ISCE3 kept about 5.6 of the 16~vCPUs busy. The scripts, records and a longer account of each comparison are in the directory \texttt{results/comparison} of the study repository.

RITSAR's error arises from its index construction. The range axis is built with \texttt{linspace}, which makes the spacing too large by one part in $N_{\rm fft}-1$, and the frequency axis assumes an even sample count, which the 14,399 samples of the collection do not satisfy. The error of \texttt{bpBasic} falls by 12 to 15~dB per doubling of its FFT length, as the second-order error of linear interpolation predicts. At $2^{17}$ it stays 15 to 17~dB above FastSAR's exact backprojection with linear interpolation of 8-times oversampled profiles (region errors in the study records); a plausible cause, not tested, is that \texttt{bpBasic} references the phase of its profiles to the lowest frequency of the band. The unmodified MatX example's error (Section~\ref{sec:oss}) arises in the same way from a float32 \texttt{linspace} for the pixel coordinates; the example's own validation compares its float32 modes with its float64 mode, which uses the same float32 coordinates, so it cannot detect the error.

ISCE3's amplitude correlates with the reference at 0.999 or more; most of its complex difference is a linear phase ramp common to all patches, consistent with a difference in spectral centering. Once the ramp is removed, the center patch agrees to $-54.9$~dB and the off-center patches retain an unexplained residual of $-31.5$ to $-31.9$~dB.

torchbp evaluates the pixel-to-antenna distance in float32, and at the 744~km range of this collection adjacent float32 values are 6~cm apart, about two wavelengths, so its output has no correlation with the reference (0.0~dB after the fitted gain, the entry of Table~\ref{tab:oss}). A NumPy reimplementation of the same range formula in float64 forms a 128 by 128~pixel crop of the lock region with an error of $-28.3$~dB, the level expected from linear interpolation of 2.28-times oversampled profiles. The same reimplementation in float32 reproduces the failure, so the cause is the float32 range.

torchbp's factorized backprojection required more than 23~GiB, which exceeds the 22.0~GiB that PyTorch reports as the L4's capacity. GRDL's factorized backprojection, which its documentation describes as a stripmap algorithm, ran on the CPU in 77.5~s. Its image of the spotlight collection is smeared in cross-range, and no orientation of its grid registers to the reference (amplitude correlation 0.22 or less).

The AFRL toolbox also contains a polar format, \texttt{pfa\_mem}, run unmodified in Octave on the same instance (with a compatibility version of \texttt{cross} for mixed row and column inputs), and GRDL's polar format ran on the same instance. Neither image is corrected for the planar-wavefront displacement. Each region was therefore registered and resampled onto the reference pixels before scoring by amplitude statistics and coherence, as for the vendor image of Appendix~\ref{app:sicd} (Table~\ref{tab:osspfa}); part of their coherence loss comes from that resampling.

\begin{table}[tbp]
\centering
\caption{FastSAR's polar format is closer to the reference than the two open-source polar formats and faster than both on the CPU instance. Correlation of the log amplitude with the reference and mean coherence in 5 by 5~pixel windows over the three Panama regions, and the time per full image.}
\label{tab:osspfa}
\small
\begin{tabular}{lrrr}
\toprule
Implementation & Log-amplitude correlation & Coherence & Time per full image \\
\midrule
AFRL toolbox \texttt{pfa\_mem} (Octave) & 0.80 to 0.89 & 0.76 to 0.88 & 44.5~s \\
GRDL \texttt{PolarFormatAlgorithm} & 0.78 to 0.93 & 0.83 to 0.94 & 24.5~s \\
FastSAR polar format, float32 & 0.996 to 0.999 & 0.999 & 9.0~s \\
\bottomrule
\end{tabular}
\end{table}

\section{Measurement protocol in detail}
\label{app:protocol}

This section gives the software versions, the timing rules, the power basis, the image-quality statistics and grades (Table~\ref{tab:prec}) and the convergence of the reference.

\paragraph{Software and timing rules.} The software comprised FastSAR release 0.1.7 (tag \texttt{v0.1.7}), JAX 0.11.2~\cite{jax2018}, CUDA 12 with CuPy 14.2.0~\cite{okuta2017}, and Numba 0.68.0~\cite{lam2015} for the reference. The TPU read rates of Appendix~\ref{app:tpu} and the two exact-backprojection region errors against the 64-times computation (Table~\ref{tab:oss}, Figure~\ref{fig:teaser}) were measured at an earlier tag, \texttt{rel-rc6-20261009}, whose exact kernels form the same images as the release. The region times behind the full-image estimate of Section~\ref{sec:oss} were recorded with the release, together with the full-image times. Every configuration ran in its own process through the public interface. Where several implementation variants give errors within 0.1~dB of each other, the fastest is the one timed.

\paragraph{Reproducibility of the records.} The images of the release are bitwise identical across calls, processes and instances (checksums of every record image are in the repository). The reason is that the factorized CUDA kernel rotates each child from the raw samples rather than in place (Appendix~\ref{app:kernels}), so no image depends on the device memory free at the call. That determinism costs 4 to 8\% of the factorized time on the H100, 8 to 10\% on the A100 and under 5\% on the L4.

Timing varies between instances of the same type by up to 7\% on the L4, the largest instance effect observed and above the 3\% timing resolution; the tables carry the first instance measured. The shorter of two warm calls is reported because host preparation of the largest phase histories is occasionally many times slower on one call.

\paragraph{Time and cost.} For the factorized algorithm the timed interval of Section~\ref{sec:protocol} includes the per-collection host work (the float64 geometry of all three levels for the CUDA and C++ kernels and of the first two for the TPU kernels). Timing followed the two-warm-call rule of Section~\ref{sec:protocol}, except for the Capella mosaics, each timed as a single mosaic after a first. The first warm call was up to 90\% slower than the second where one-time work fell on it (90th percentile of the difference 56\%); this happened on every GPU in the float16, exact and polar-format configurations, by 0.2 to 0.8~s, and in no float32 factorized record. On the CPU and the L4 the reported time agrees with the timed loop's time per image within 3\%. The three scored regions of Figure~\ref{fig:teaser} are pooled with weights equal to their reference energy; the A100 and H100, which run the L4's CUDA code, carry the L4's region errors there, their whole-image errors agreeing with the L4's within 0.002~dB.

The loop ran until at least four images had been formed and at least 60~s had elapsed. It formed images in sequence on the CPU and the GPUs and, on the TPUs, staged the next phase history on a worker thread (\texttt{ImageFormer.stage}, as in the mosaics) while the current image formed; on the CPU, exact backprojection and the JAX program were timed without a loop. The factorized kernels were tuned against the bounds of Table~\ref{tab:bounds}; the polar-format kernels were not profiled against bounds. The L4 was hosted by a g2-standard-4 instance for FastSAR on the Umbra collections, a g2-standard-8 for the open-source implementations (whose L4 costs in Figure~\ref{fig:teaser} use the g2-standard-4 price of Table~\ref{tab:platforms}) and a g2-standard-32 for the Capella collections. The process's CPU time was recorded during every run.

\paragraph{Power.} The draw of the L4, A100 and H100 was sampled from \texttt{nvidia-smi} twice a second during the timed loops, and Table~\ref{tab:power} gives the means. The TPU virtual machines and the CPU instance expose no power counters. The TPU figures are ranges from design-power figures quoted by third parties (120 to 200~W for the v5e, 200 to 350~W for the v6e~\cite{introl2025,gpuadvisor2025}), since the announcement of the v6e~\cite{google2024trillium} gives no power figure, assuming the chip ran at that power throughout; the kWh column of Table~\ref{tab:cost} takes the midpoint. The CPU figure is the pro-rata share of the 8 cores (25~W of the 400~W package) times their occupancy, the busy fraction of the 16 hardware threads doubled and capped at one. Hosts, memory and the other components of each instance are excluded, and the TPU and CPU figures are marked as estimates. The table uses the single-image time; the kWh column of Table~\ref{tab:cost} uses the timed loop's per-image time, shorter on the TPUs and within 3\% elsewhere.

\begin{table}[tbp]
\centering
\caption{The L4's factorized configurations use the least energy per image in the table, 0.17 to 0.20~kJ against at least 0.28~kJ on the other devices; the TPU figures are estimates from design-power bands that span a factor of 1.7 to 1.8. Energy per Panama image for the processor alone over the single-image time: the power basis of each device and the energy of its configurations in Table~\ref{tab:cost}. Derivation in Appendix~\ref{app:protocol}.}
\label{tab:power}
\small
\setlength{\tabcolsep}{4pt}
\resizebox{\textwidth}{!}{\begin{tabular}{lllrrr}
\toprule
Device & Configuration & Basis & Power (W) & Time (s) & Energy (kJ) \\
\midrule
L4 & Exact BP, cubic interpolation, $8\times$ oversampled & measured & 72 & 16.8 & 1.20 \\
L4 & Factorized BP, float16 throughout, CUDA kernels & measured & 68 & 2.5 & 0.17 \\
L4 & Factorized BP, float32, CUDA kernels & measured & 68 & 3.0 & 0.20 \\
TPU v5e & Factorized BP, single-pass bfloat16, TPU kernels & third-party design power & 120 to 200 & 3.5 & 0.4 to 0.7 \\
TPU v5e & Factorized BP, three-pass bfloat16, TPU kernels & third-party design power & 120 to 200 & 5.0 & 0.6 to 1.0 \\
TPU v6e & Factorized BP, single-pass bfloat16, TPU kernels & third-party design power & 200 to 350 & 2.5 & 0.5 to 0.9 \\
TPU v6e & Factorized BP, three-pass bfloat16, TPU kernels & third-party design power & 200 to 350 & 2.9 & 0.6 to 1.0 \\
CPU & Factorized BP, float32, C++ kernels & pro-rata estimate & 25 at 100\% & 11.2 & 0.28 \\
\bottomrule
\end{tabular}}
\end{table}

\paragraph{Image-quality statistics and grades.} The grades use the thresholds of Section~\ref{sec:quality}, 2.3$^\circ$ of phase and 0.1~dB of amplitude at the 99th percentile, over two populations: the phase error over the brightest 1\% of the reference pixels of each region, where persistent scatterers are measured, and the amplitude ratio over the brighter half of the pixels, since radiometry concerns distributed targets. Grade~1 meets both requirements, grade~2 the phase requirement only, and grade~3 neither. Table~\ref{tab:prec} lists both statistics and the resulting grade for every Panama configuration.

The bright-scatterer statistics are computed on the four 512 by 512~pixel Panama regions after the same fitted gain (script \texttt{oss/bright\_scatterer\_errors.py}, record \texttt{results/quality/\allowbreak bright\_scatterers.json}): at the 99th percentile the phase error is at most 0.1$^\circ$ for the float32-class, float16-throughout and exact configurations, 0.2$^\circ$ for single-pass bfloat16 and 4.8$^\circ$ for polar format, with maxima of 0.5, 0.5 and 11.6$^\circ$. The gain fitted per algorithm family is a complex gain for the backprojections and a real one for polar format, whose phase differs from the reference by a smooth screen before the final resampling.

Six statistics are computed for each image. The first, the error, is the energy of the difference relative to the energy of the reference (dB); this is the figure quoted when a single number is given. The second is the largest single-pixel difference relative to the brightest pixel (dB). The third is the complex coherence between the two images over a 5 by 5 window, reported as the mean, the 0.1 percentile and the minimum. The fourth and fifth are the amplitude ratio and the phase difference over the brighter half of the pixels, reported as the 99th percentile and the maximum. The sixth is the error in four rings by distance from the scene center. All six, for every configuration on every collection, are in the tables of the study repository (\texttt{paper/supplement\_tables.pdf}), whose configuration labels are those of Table~\ref{tab:arith}.

\begin{table}[htbp]
\centering
\caption{Every float32-class, float16-throughout and exact image is grade~1; single-pass bfloat16 fails only the amplitude requirement (0.16~dB at the 99th percentile) and polar format fails both. Image-quality statistics of the Panama images against the reference (definitions in Appendix~\ref{app:protocol}): error, largest pixel difference relative to the brightest pixel (dB), 0.1 percentile and minimum of the 5 by 5 coherence, the amplitude ratio at the 99th percentile over the brighter half of the pixels (dB), the phase error at the 99th percentile and its maximum over the brightest 1\% of the reference pixels of each region (degrees), and the resulting grade; the 16-times oversampled exact row uses the reference's interpolation.}
\label{tab:prec}
\small
\setlength{\tabcolsep}{4pt}
\resizebox{\textwidth}{!}{\begin{tabular}{llrrrrrrrr}
\toprule
Device & Configuration & Error (dB) & Max (dB) & Coh. p0.1 & Coh. min & Amp. p99 (dB) & Phase p99 ($^\circ$) & Phase max ($^\circ$) & Grade \\
\midrule
L4 & Exact BP, cubic interpolation, $8\times$ oversampled & $-67.4$ & $-72.1$ & 1.0000 & 0.998 & 0.01 & 0.0 & 0.2 & 1 \\
CPU & Exact BP, cubic interpolation, $8\times$ oversampled & $-67.4$ & $-72.1$ & 1.0000 & 0.998 & 0.01 & 0.0 & 0.2 & 1 \\
L4 & Exact BP, linear interpolation, $8\times$ oversampled & $-56.8$ & $-62.2$ & 0.9998 & 0.922 & 0.03 & 0.1 & 0.5 & 1 \\
CPU & Exact BP, linear interpolation, $8\times$ oversampled & $-56.9$ & $-62.2$ & 0.9998 & 0.923 & 0.03 & 0.1 & 0.5 & 1 \\
L4 & Exact BP, linear interpolation, $16\times$ oversampled & $-75.2$ & $-88.9$ & 1.0000 & 1.000 & 0.01 & 0.0 & 0.0 & 1 \\
L4 & Factorized BP, float32, CUDA kernels & $-60.0$ & $-64.9$ & 0.9999 & 0.931 & 0.02 & 0.0 & 0.2 & 1 \\
L4 & Factorized BP, float16 throughout, CUDA kernels & $-58.9$ & $-64.9$ & 0.9999 & 0.931 & 0.03 & 0.1 & 0.2 & 1 \\
CPU & Factorized BP, float32, C++ kernels & $-60.0$ & $-65.0$ & 0.9999 & 0.931 & 0.02 & 0.0 & 0.2 & 1 \\
CPU & Factorized BP, float32, JAX program & $-58.3$ & $-64.8$ & 0.9999 & 0.931 & 0.03 & 0.1 & 0.2 & 1 \\
TPU v6e & Factorized BP, three-pass bfloat16, TPU kernels & $-57.8$ & $-64.5$ & 0.9999 & 0.931 & 0.03 & 0.1 & 0.2 & 1 \\
TPU v6e & Factorized BP, single-pass bfloat16, TPU kernels & $-46.2$ & $-64.8$ & 0.9981 & 0.674 & 0.16 & 0.2 & 0.4 & 2 \\
TPU v5e & Factorized BP, three-pass bfloat16, TPU kernels & $-57.9$ & $-64.8$ & 0.9999 & 0.931 & 0.03 & 0.1 & 0.2 & 1 \\
TPU v5e & Factorized BP, single-pass bfloat16, TPU kernels & $-46.2$ & $-65.0$ & 0.9981 & 0.674 & 0.16 & 0.2 & 0.5 & 2 \\
L4 & Polar format, float32, gather & $-32.1$ & $-40.3$ & 0.9857 & 0.674 & 0.69 & 4.8 & 11.6 & 3 \\
TPU v6e & Polar format, float32, gather & $-32.1$ & $-40.3$ & 0.9857 & 0.680 & 0.69 & 4.8 & 11.6 & 3 \\
TPU v5e & Polar format, float32, gather & $-32.1$ & $-40.3$ & 0.9857 & 0.680 & 0.69 & 4.8 & 11.6 & 3 \\
CPU & Polar format, float32, gather & $-32.1$ & $-40.3$ & 0.9857 & 0.674 & 0.69 & 4.8 & 11.6 & 3 \\
\bottomrule
\end{tabular}}
\end{table}

\paragraph{CPU details.} On the CPU the factorized float32 row comes from the C++ kernels of Appendix~\ref{app:kernels}, exact backprojection from its C++ kernel and polar format from its C++ gather and resampling kernels. The reference was formed on a separate n2-highmem-48 instance (48~vCPUs, \$3.144 per hour on demand) in 1,892~s per Panama image (about \$1,652 per thousand images) and is not part of the cost comparison.

\paragraph{Reference convergence and exact backprojection.} The reference's convergence was measured on a 1024 by 1024 crop by forming the same crop with 8-, 16- and 32-times oversampled profiles. The 8-times image differs from the 32-times image by $-55.5$~dB and the 16-times image by $-69.2$~dB. The 16-times reference was adopted because the convergence error at 8-times oversampling lies within the range of the errors under study. On the lock, port, and ship regions the same float64 code with 64-times oversampled profiles (the 64-times computation of Section~\ref{sec:protocol}) differs from the reference by $-67.3$ to $-68.8$~dB, an estimate of the reference's own error in these regions. Against the 64-times computation the region errors of the float32-class factorized images are $-57.2$ to $-62.6$~dB on the CPU, $-57.1$ to $-62.6$~dB on the L4, and $-56.0$ to $-58.3$ and $-56.0$ to $-59.0$~dB with three-pass products on the v6e and v5e; these are the ranges summarized in Section~\ref{sec:concl}.

FastSAR's exact backprojection with linear interpolation of 16-times oversampled profiles, the interpolation of the reference, measures $-75.2$~dB against it; the shared interpolation error cancels, and the figure indicates only agreement between the two implementations. With linear interpolation the error is set by the profile oversampling, $-56.8$~dB at 8-times oversampling. Against the 64-times computation, cubic interpolation of 8-times oversampled profiles, the library default, gives $-77.5$ to $-81.7$~dB in the three regions, and cubic interpolation of 4-times oversampled profiles $-59.8$ to $-69.6$~dB. Linear interpolation of oversampled profiles is the scheme of the open-source implementations (Table~\ref{tab:oss} gives each one's oversampling).

On 512 by 512~pixel grids at the same regions, cubic interpolation at 16-times and linear at 64-times oversampling are within 0.3~dB of cubic at 8-times oversampling, so more oversampling does not lower the error; the float32 arithmetic and the third-order range expansion of the kernel are the likely source of this floor. With linear interpolation of 16-times oversampled profiles the L4 takes 13.8~s per image, about \$2.68 per thousand images.

\paragraph{Limitations.} Interferometric and change-detection products, the AVX-512 bfloat16 path of the CPU and a separable polar-format resampling were not evaluated, and the Taylor-weighting hypothesis of Appendix~\ref{app:pfa} was not tested. The improvement still identified in the factorized kernels is about 12\% on the v6e; the GPU kernels were profiled stage by stage only on the L4, and on an H100 only the first level was timed (profiles in the repository); no improvement was quantified on the CPU, whose stages run at 1.05 to 1.4 times their bounds (Appendix~\ref{app:kernels}). The factorized plan (32-pixel final tiles, three levels, a passband fraction of 0.4 and 70~dB filter attenuation) kept its defaults on every device, and a different factorization would change the bounds themselves by an amount not measured here. The prices are on-demand prices, with no spot or committed-use discount, for instances whose hosts differ in size (24 and 44~vCPUs for the TPUs, the smallest offered, 4~vCPUs, for the L4). TPU7x instances were outside the project's cloud quota.

\section{Ground truth of the Panama regions}
\label{app:truth}

\begin{figure}[tbp]
\centering
\includegraphics[width=0.96\textwidth]{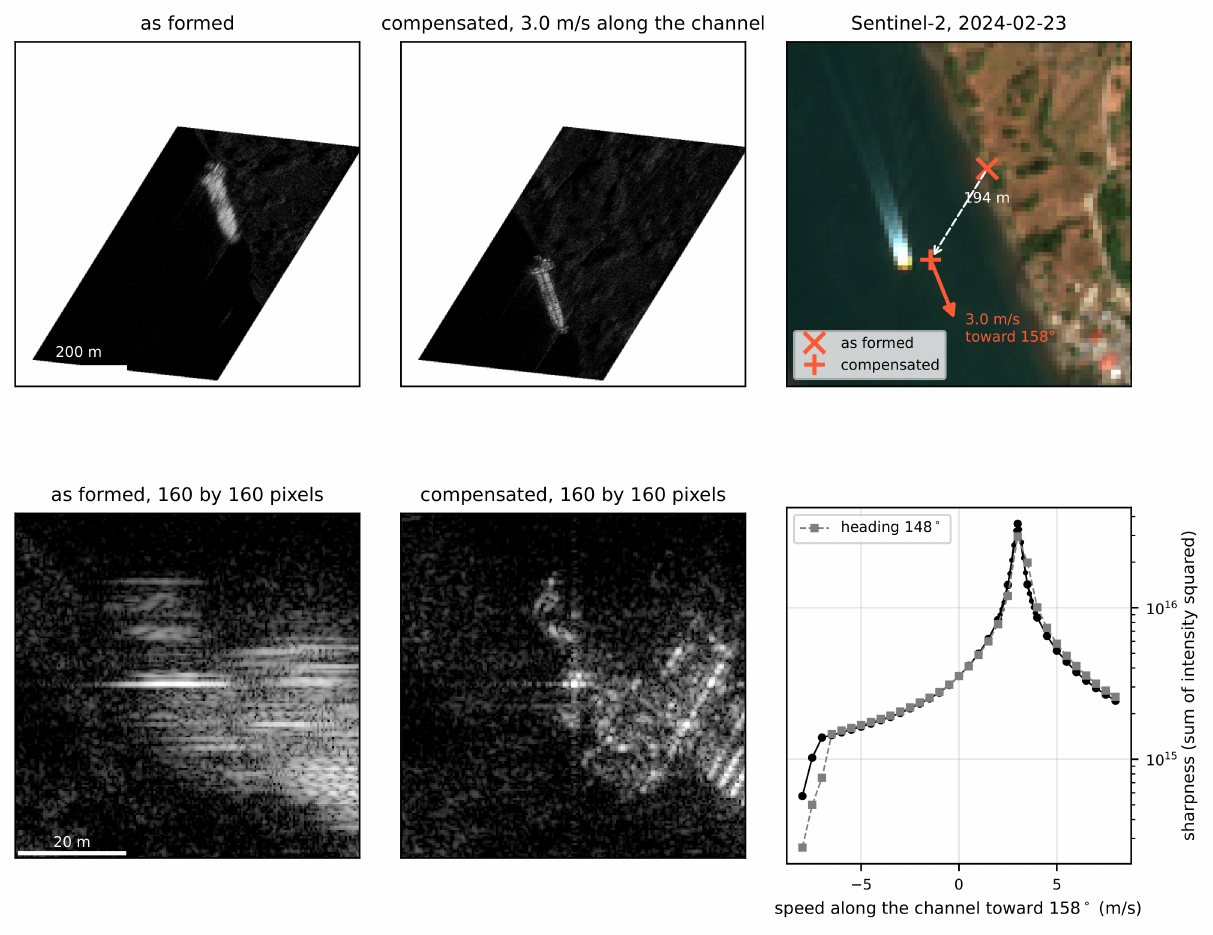}
\caption{Backprojecting with the track shifted for a 3.0~m/s ship focuses the target 10.6~dB brighter and moves it 194~m into the navigation channel. Top: the tile around the ship as formed and after backprojection with the antenna track shifted for a ship moving at 3.0~m/s along the channel, both resampled north up over a 760~m window (200~m bar), and the two positions on the Sentinel-2 image of February 2024 with the displacement (dashed) and the estimated velocity, 3.0~m/s toward $158^\circ$, drawn at the compensated position. Bottom: 160 by 160~pixel windows around the ship in image coordinates (20~m bar), and the sharpness of the tile against the assumed speed for the channel heading of $158^\circ$ (coarse and fine sweeps) and for $148^\circ$.}
\label{fig:ship}
\end{figure}

The identity of the Panama regions of Figure~\ref{fig:zoom} was checked against Copernicus Sentinel-2 imagery~\cite{sentinel2}, and the ship region needed a moving-target explanation, because its bright target lies on the bank rather than in the channel (Figure~\ref{fig:ship}). Each pixel of the images formed here was located on the ground from the vendor's SICD header. The pixel's position in the slant plane follows from the scene center point and the row and column unit vectors. The ground point is the intersection of the line through it along the slant plane's normal with the WGS84 ellipsoid at the scene height. The image corners so computed agree with the corner coordinates in the SICD header to within 15~m.

The nearest usable Sentinel-2 acquisition is from July 23, 2023, five days after the collection, at 10~m resolution; a dry-season acquisition of February 23, 2024 is shown as well because clouds cover part of the earlier one. The lock chambers, the quays and container yard of the terminal, and the road beside the vegetation region coincide in the two sensors.

The bright target of the ship region, about 180~m long and close to the eastern bank, does not appear in the vendor's geocoded products of six other passes of this repeat-collection site. Four of those passes are three, four, five and seven days after this collection and two are in February 2024. The pass of July 25, 2023 (02:30 UTC) repeats the geometry of this collection to within $0.3^\circ$ (azimuth $122.4^\circ$ versus $122.1^\circ$, grazing angle $42.6^\circ$ in both, left-looking). The other passes span azimuths from $-72^\circ$ to $223^\circ$ and grazing angles from $40^\circ$ to $63^\circ$, so an aspect-dependent return from a fixed object is unlikely.

The pass of July 23, 2023 (14:47 UTC) precedes the Sentinel-2 image of that day by 64~min, and neither shows anything at the position. It is taken to be a ship in the navigation channel; the Sentinel-2 image of February 2024 shows a different ship nearby. The vendor's geocoded product of this collection places the target inside the outlined crop as well.

The ship's apparent position on the eastern bank, inland of the channel, is consistent with its motion during the aperture (Figure~\ref{fig:ship}). A stationary-scene processor places a target with line-of-sight velocity $v_r$ at the position whose Doppler history matches, displaced along azimuth by about $R v_r / V$ (here about 100~m per m/s), and the along-track velocity defocuses it~\cite{raney1971}. Exact backprojection takes the antenna position of every pulse as data, so a target moving at constant velocity $\mathbf{v}$ is formed exactly by backprojecting with the track shifted by $-\mathbf{v} t$. The target is then focused and placed at its position at the aperture center.

The ship was assumed to follow the channel axis (bearing $158^\circ$, from the Sentinel-2 water mask) and its speed was swept from $-8$ to $8$~m/s in a tile around it (Figure~\ref{fig:ship}). The sharpness of the tile, the sum of the squared intensities, is single-peaked at $3.0$~m/s (5.8~knots toward the Pacific). At that speed the peak intensity is 10.6~dB above the stationary-scene image and the ship has moved 469~pixels (194~m) along azimuth into the navigation channel; the speed estimate uses only the sharpness of the tile, and the position of the compensated target was not used in the fit. A heading $10^\circ$ farther east ($148^\circ$) gives the same speed with an 18\% lower sharpness peak and a compensated position 22~m away. The images compared in Figure~\ref{fig:zoom} are the stationary-scene images, in which the ship appears at the same displaced position under every algorithm.

\section{Melbourne, Iowa and the Capella collections}
\label{app:modes}
\label{app:stats}

On Melbourne and Iowa the factorized errors lie within 1.2~dB of Panama's on every device and the L4 stays the cheapest device for every algorithm (Table~\ref{tab:costother}); on the five Capella collections the H100 is the fastest device and the L4 the cheapest wherever its memory holds the default first-level groups (Table~\ref{tab:modes}). The Melbourne and Iowa rows repeat the Panama configurations of Table~\ref{tab:cost} on smaller, less squinted images, and their times follow the pixel count.

\begin{table}[tbp]
\centering
\caption{On Melbourne and Iowa every factorized error is within 1.2~dB of Panama's and the L4 is again the cheapest device for every algorithm. Time, throughput, cost, energy and whole-image error of the configurations of Table~\ref{tab:cost} except exact backprojection on the CPU, with the arithmetic labels of that table shortened (the kernels are those of Table~\ref{tab:arith}); bold marks the fastest configuration on each collection.}
\label{tab:costother}
\small
\setlength{\tabcolsep}{3pt}
\fitwidth{\begin{tabular}{lllrrrrr}
\toprule
Device & Algorithm & Arithmetic & Time (s) & Images/h & \$/1000 & kWh/1000 & Error (dB) \\
\midrule
\multicolumn{8}{l}{\emph{Melbourne}} \\
L4 & Factorized BP & float16 throughout & 2.6 & 1,375 & 0.51 & 0.05 & $-58.5$ \\
L4 & Factorized BP & float32 & 3.0 & 1,172 & 0.60 & 0.06 & $-59.4$ \\
L4 & Polar format & float32 & 1.5 & 2,389 & 0.30 & 0.02 & $-33.5$ \\
L4 & Exact BP & cubic, $8\times$ & 17.0 & 211 & 3.36 & 0.34 & $-67.2$ \\
L4 & Exact BP & linear, $8\times$ & 11.1 & 327 & 2.16 & 0.22 & $-56.8$ \\
A100 & Factorized BP & float16 throughout & 1.1 & 3,209 & 1.56 & 0.08 & $-58.5$ \\
A100 & Factorized BP & float32 & 1.2 & 2,977 & 1.69 & 0.10 & $-59.4$ \\
A100 & Polar format & float32 & 0.9 & 3,835 & 1.31 & 0.03 & $-33.5$ \\
A100 & Exact BP & cubic, $8\times$ & 8.4 & 429 & 11.70 & 0.88 & $-67.2$ \\
\textbf{H100} & \textbf{Factorized BP} & \textbf{float16 throughout} & \textbf{0.6} & \textbf{6,304} & \textbf{1.74} & \textbf{0.07} & $\mathbf{-58.5}$ \\
\textbf{H100} & \textbf{Factorized BP} & \textbf{float32} & \textbf{0.6} & \textbf{5,808} & \textbf{1.89} & \textbf{0.09} & $\mathbf{-59.4}$ \\
\textbf{H100} & \textbf{Polar format} & \textbf{float32} & \textbf{0.6} & \textbf{6,186} & \textbf{1.77} & \textbf{0.03} & $\mathbf{-33.5}$ \\
H100 & Exact BP & cubic, $8\times$ & 4.5 & 793 & 13.84 & 0.72 & $-67.2$ \\
TPU v5e & Factorized BP & single-pass bfloat16 & 3.6 & 1,463 & 0.82 & 0.11 & $-46.2$ \\
TPU v5e & Factorized BP & three-pass bfloat16 & 5.2 & 866 & 1.39 & 0.18 & $-57.9$ \\
TPU v5e & Polar format & float32 & 68.2 & 53 & 22.73 & 3.03 & $-33.5$ \\
TPU v6e & Factorized BP & single-pass bfloat16 & 2.5 & 2,380 & 1.13 & 0.12 & $-46.2$ \\
TPU v6e & Factorized BP & three-pass bfloat16 & 2.9 & 1,803 & 1.50 & 0.15 & $-57.7$ \\
TPU v6e & Polar format & float32 & 106.4 & 34 & 79.81 & 8.13 & $-33.5$ \\
CPU & Factorized BP & float32 & 11.8 & 305 & 3.16 & 0.08 & $-59.5$ \\
CPU & Polar format & float32 & 9.0 & 398 & 2.42 & 0.06 & $-33.5$ \\
\midrule
\multicolumn{8}{l}{\emph{Iowa farmland}} \\
L4 & Factorized BP & float16 throughout & 1.6 & 2,180 & 0.32 & 0.03 & $-59.1$ \\
L4 & Factorized BP & float32 & 1.9 & 1,867 & 0.38 & 0.04 & $-60.2$ \\
L4 & Polar format & float32 & 1.1 & 3,351 & 0.21 & 0.02 & $-35.4$ \\
L4 & Exact BP & cubic, $8\times$ & 10.8 & 336 & 2.11 & 0.21 & $-70.7$ \\
L4 & Exact BP & linear, $8\times$ & 6.7 & 543 & 1.30 & 0.13 & $-60.4$ \\
A100 & Factorized BP & float16 throughout & 0.7 & 4,984 & 1.01 & 0.05 & $-59.1$ \\
A100 & Factorized BP & float32 & 0.8 & 4,682 & 1.07 & 0.06 & $-60.2$ \\
A100 & Polar format & float32 & 0.7 & 5,403 & 0.93 & 0.02 & $-35.4$ \\
A100 & Exact BP & cubic, $8\times$ & 5.0 & 714 & 7.03 & 0.52 & $-70.7$ \\
\textbf{H100} & \textbf{Factorized BP} & \textbf{float16 throughout} & \textbf{0.4} & \textbf{9,818} & \textbf{1.12} & \textbf{0.05} & $\mathbf{-59.1}$ \\
\textbf{H100} & \textbf{Factorized BP} & \textbf{float32} & \textbf{0.4} & \textbf{9,076} & \textbf{1.21} & \textbf{0.05} & $\mathbf{-60.2}$ \\
\textbf{H100} & \textbf{Polar format} & \textbf{float32} & \textbf{0.4} & \textbf{8,959} & \textbf{1.22} & \textbf{0.02} & $\mathbf{-35.4}$ \\
H100 & Exact BP & cubic, $8\times$ & 2.9 & 1,254 & 8.75 & 0.43 & $-70.7$ \\
TPU v5e & Factorized BP & single-pass bfloat16 & 2.4 & 2,176 & 0.55 & 0.07 & $-46.3$ \\
TPU v5e & Factorized BP & three-pass bfloat16 & 3.4 & 1,340 & 0.90 & 0.12 & $-58.8$ \\
TPU v5e & Polar format & float32 & 48.3 & 75 & 16.09 & 2.15 & $-35.4$ \\
TPU v6e & Factorized BP & single-pass bfloat16 & 1.7 & 3,371 & 0.80 & 0.08 & $-46.3$ \\
TPU v6e & Factorized BP & three-pass bfloat16 & 2.1 & 2,546 & 1.06 & 0.11 & $-58.9$ \\
TPU v6e & Polar format & float32 & 61.1 & 59 & 45.81 & 4.67 & $-35.4$ \\
CPU & Factorized BP & float32 & 7.0 & 513 & 1.88 & 0.05 & $-60.2$ \\
CPU & Polar format & float32 & 6.7 & 539 & 1.79 & 0.05 & $-35.4$ \\
\bottomrule
\end{tabular}}
\end{table}

On the five Capella collections (Table~\ref{tab:modes}), FastSAR's worst-region error over the five devices lies between $-59.1$ and $-61.5$~dB. On the sliding spotlight, however, FastSAR's image matches the vendor's only near the scene center, because of the range-ambiguity convention of that file (Appendix~\ref{app:sicd}).

On the CPU instance, ISCE3's backprojection would take an estimated 41 to 3{,}600 times as long as FastSAR. Against the reference formed with the CPHD positions, ISCE3's worst-region error is $-25.8$ to $-50.6$~dB, about 10 to 34~dB above FastSAR's. Because ISCE3's receive model differs from FastSAR's (Appendix~\ref{app:algorithms}), it was also scored against a reference formed with its own receive model. There its worst-region error is 4.0 to 16.6~dB below FastSAR's worst error over the five devices on the 2024 spotlight, the 2025 spotlight and the 2021 stripmap. On the 2022 sliding spotlight and the 2025 stripmap it is 0.2 and 1.4~dB above.

Among the FastSAR devices the H100 has the shortest time on all five collections, the v6e's time being 1.5 to 4.1 times its own (Table~\ref{tab:modes}; the ratios in this section are computed from the unrounded times and costs of the records). On the two stripmaps the L4 is cheapest, 16\% below the v6e on the 2025 collection and 27\% on the 2021 one, and on the sliding spotlight the v6e is 2\% below the L4, within the timing resolution (\$25 per thousand images on the v6e against \$26 on the L4).

For the two spotlights the L4 forms the image with smaller first-level groups, because the default group needs more memory than the L4 has free (note to Table~\ref{tab:modes}), and takes 2.9 and 3.6 times as long as the A100. The A100 is cheapest on the 2025 spotlight, at 0.8 times the L4's cost; on the 2024 one the L4 and the A100 cost the same within 2\%. The three mosaics (the sliding spotlight and the two stripmaps) show the A100 up to 1.8 times faster than the L4 at 1.6 to 3.1 times its cost, and the H100 at 1.8 to 3.7 times the L4's cost; the rounded costs of Table~\ref{tab:modes} give 1.5 and 1.7 on the sliding spotlight.

\begin{table}[htbp]
\centering
\caption{The H100 is fastest on every Capella collection; the L4 is cheapest on both stripmaps and the A100 on the 2025 spotlight, whose first-level groups exceed the L4's memory. Top: pulses, megapixels (Mpx), and FastSAR's time per image and cost per thousand images at float32-class accuracy on five devices. Bottom: FastSAR's worst region error over the devices, the estimated time of ISCE3 on the CPU instance with its worst-region error under its own receive model and under the CPHD positions, and the algorithm and time of GRDL on the CPU instance (conventions in the note).}
\label{tab:modes}
\footnotesize
\setlength{\tabcolsep}{3.5pt}
\fitwidth{\begin{tabular}{lrrrrrrrrrrrr}
\toprule
 & & & \multicolumn{5}{c}{FastSAR time (s)} & \multicolumn{5}{c}{FastSAR cost (\$/1000)} \\
\cmidrule(lr){4-8}\cmidrule(lr){9-13}
Collection & Pulses & Mpx & CPU & L4 & A100 & H100 & v6e & CPU & L4 & A100 & H100 & v6e \\
\midrule
Spotlight, 2024 & 74{,}203 & 258 & 53.7 & 15.5* & 5.3 & 2.7 & 10.9 & 14 & 7.5* & 7.4 & 8.3 & 8.2 \\
Spotlight, 2025 & 54{,}267 & 255 & 46.0 & 16.3* & 4.5 & 2.3 & 9.6 & 12 & 7.9* & 6.3 & 7.1 & 7.2 \\
Sliding spotlight, 2022 & 123{,}799 & 72 & 254 & 53.4 & 29.0 & 14.9 & 33.8 & 68 & 26 & 40 & 45 & 25 \\
Stripmap, 2021 & 39{,}898 & 83 & 44.8 & 9.5 & 10.3 & 5.6 & 8.4 & 12 & 4.6 & 14 & 17 & 6.3 \\
Stripmap, 2025 & 237{,}720 & 1{,}132 & 842 & 195 & 136 & 72.3 & 149 & 225 & 94 & 189 & 220 & 112 \\
\bottomrule
\end{tabular}}
\\[1.5ex]
\fitwidth{\begin{tabular}{lrrrrlr}
\toprule
 & FastSAR & \multicolumn{3}{c}{ISCE3} & \multicolumn{2}{c}{GRDL} \\
\cmidrule(lr){3-5}\cmidrule(lr){6-7}
Collection & error (dB) & time (h) & own model (dB) & CPHD (dB) & alg. & time (s) \\
\midrule
Spotlight, 2024 & $-59.1$ & 54.1 & $-63.1$ & $-25.8$ & PFA & 99 \\
Spotlight, 2025 & $-59.2$ & 41.6 & $-75.7$ & $-36.3$ & PFA & 70 \\
Sliding spotlight, 2022 & $-61.5$ & 2.9 & $-61.3$ & $-38.6$ & -- & -- \\
Stripmap, 2021 & $-60.2$ & 2.3 & $-64.8$ & $-50.6$ & RDA & 28 \\
Stripmap, 2025 & $-60.6$ & 48.2 & $-59.2$ & $-26.9$ & SMPFA & 550 \\
\bottomrule
\end{tabular}}
\par\vspace{3pt}\parbox{\textwidth}{\footnotesize FastSAR's time per image is host memory to host memory (Section~\ref{sec:protocol}) and its cost the hourly price times the single-image time. The spotlights are timed as the shorter of two warm calls, which agree within 2\%, the mosaics as a single mosaic after a first one. The L4 ran on a g2-standard-32 instance at \$1.734 per hour, the A100 (80~GB) on an a2-ultragpu-1g at \$5.02 and the H100 (80~GB) on an a3-highgpu-1g at \$10.97. Rows marked with an asterisk ran with smaller first-level groups (with a \texttt{MemoryWarning}), because full speed needs more than the 23.0~GB (21.4~GiB) free on the L4: 35.8~GB for the 2024 spotlight and 25.4~GB for the 2025 one. FastSAR's error is the worst of three 512 by 512~pixel regions against the float64 reference over the five devices; ISCE3's whole-image time is estimated from the three regions on the CPU instance. GRDL's slant-plane images are not scored against the reference.}
\end{table}

\end{document}